\documentstyle[epsfig]{cupconf}
\begin{document}
%
\let\ov=\over
\let\lbar=\l
\let\l=\left
\let\r=\right
\def \der#1#2{{\partial{#1}\over\partial{#2}}}
\def \dder#1#2{{\partial^2{#1}\over\partial{#2}^2}}
\def\N{{I\!\!N}}
\def\be{\begin{equation}}
\def\ee{\end{equation}}
\def\bea{\begin{eqnarray}}
\def\eea{\end{eqnarray}}
\def\beau{\begin{eqnarray*}}
\def\eeau{\end{eqnarray*}}
\def\ms{\langle S \rangle}
\def\n2{\langle N^2 \rangle}
\def\sn2{\sqrt{\langle N^2 \rangle}}

%

%

\setcounter{part}{1}
\title[Cosmological sources]{The stochastic gravity-wave background: sources and detection}
\author[B. Allen]{B. Allen}
\affiliation{Department of Physics\\
University of Wisconsin - Milwaukee\\
PO Box 413\\
Milwaukee, WI 53211, USA\\
\vskip 5pt
email: ballen@dirac.phys.uwm.edu}


\setcounter{page}{1}
\maketitle
\vskip -2.8in
{\small \noindent To appear in: {\sl Proceedings of the Les Houches School on Astrophysical Sources of Gravitational Waves}, eds. Jean-Alain Marck and Jean-Pierre Lasota,
to be published by Cambridge University Press 1996.}
\vskip 2.4in

\firstsection


\section{Introduction}
The design and construction of a number of new and more sensitive
detectors of gravitational radiation is currently underway.  These
include the LIGO detector being built in the United States by a joint
Caltech/MIT collaboration \cite{science92}, the VIRGO detector being
built near Pisa by an Italian/French collaboration 
(Caron {\sl et al.} 1995), the
GEO-600 detector being built in Hannover by an Anglo/German
collaboration 
(Danzmann {\sl et al.} 1995), and the TAMA-300 detector being built near
Tokyo \cite{tama300}.  There are also several resonant bar detectors
currently in operation, and several more refined bar and
interferometric detectors presently in the planning and proposal
stages.

It's not clear when the first sources will be detected -- this may
happen soon after the detectors go ``on-line" or it may require decades
of further work to increase the sensitivity of the instruments.  But
it's clear to me that eventually, when the sensitivity passes above
some threshold value, the gravity-wave detectors will find sources.

The operation of these detectors will have a major impact on the field
of gravitational physics.  For the first time, there will be a
significant amount of experimental data to be analyzed, and the ``ivory
tower" relativists will be forced to interact with a broad range of
experimenters and data analysts to extract the interesting physics from
the data stream.  Being an optimist, I think that this ``interesting
physics" will include sources of gravitational radiation which we have
not hitherto expected or even conceived of.  It promises to be an
exciting time.

There are at least three major types of ``known" sources
(Schutz 1991; Thorne 1995).  These are (1) coalescing binary systems,
composed of neutron stars and/or black holes, (2) pulsars (and other
periodic sources), and (3) supernovae (and other transient or burst
sources).  These are listed, roughly speaking, in increasing order of
``detection difficulty".  For example, the waveforms (chirps) of the
coalescing binary systems are already known to a high enough degree of
precision (Blanchet {\sl et al.} 1995; Apostolatos {\sl et al.} 1994; 
Tagoshi \& Sasaki 1994) to enable reliable detection, using the
classical signal analysis technique of matched filtering
(Dhurandhar \& Sathyaprakash 1994; 
Apostolatos 1995).  Some order-of-magnitude studies have shown that the
amount of required processing can be done with a dozen or so
high-performance desk-top workstations \cite{owen}.  The major unknown
for this type of source is their number density, which is uncertain by
more than an order of magnitude.  Hence the rate of events which occur
close enough to us to be observable can not be estimated very
precisely.  The second type of source is pulsars.  Once again, the form
of the signal is known very precisely:  it's just a sine wave in the
solar system barycenter, a coordinate system at rest with respect to
the Sun \cite{pulsars}.  In this case however, the signal processing
problem is far more difficult, because the orbital motion of the earth
around the sun and the rotational motion of the earth around its axis
modulates the pulsar frequency.  This means one must separately analyze
the signal for each of $\approx 10^{14}$ separate patches on the
celestial sphere, each of which would have a distinctive pattern of
frequency modulation.  The required processing speeds are currently
beyond the limits of even the most powerful computers.  Of course, it
may be just a question of waiting until a better search algorithm is
developed, or faster computers become available!  The third source
which I've listed, supernovae, is perhaps the most difficult to
detect.  There are two reasons.  First, we do not have precise
predictions of the gravitational waveform such an event would produce
-- and it's hard to design a data analysis algorithm to look for
something unknown.  Furthermore, these explosions are expected to be
rare events and because certain kinds of instrument noise might look
quite similar to supernovae, I chose to list them last, and categorize
them as the most difficult sources to detect.

The subject of these lectures is a fourth type of source, quite
different in character from the three listed above.  These are the
``stochastic" or ``background" sources (Michelson 1987; Christensen 1992; 
Flanagan 1993).
Roughly speaking, these are ``random" sources, typically arising from
an extremely large number of ``unresolved" independent and uncorrelated
events.  This type of background could be the result of processes that
take place very shortly after the big bang, but since we know very
little about the state of the universe at that time, it's impossible to
say with any certainty.  Such a background might also arise from
processes that take place fairly recently (say within the past several
billion years) and this more recent contribution might overwhelm the
parts of the background which contain information about the state of
the early universe.

These sources are ``unresolved" in the following sense.  If we study an
optical source, somewhere in the sky, using a telescope with a certain
angular resolution, then details of the source can be ``resolved" if
the angular resolution of the telescope is smaller than the angular
size of the features or objects being studied.  In the case of the LIGO
experiment, and similar detectors, the angular size of the antenna
pattern is of order $90^\circ$.  Hence almost any source is
``unresolved" in that it makes a significant contribution to the
detector output for almost any orientation of the detector and the
source.  When many such sources are present, even if they are
pointlike, the resulting signal has a stochastic nature.

My motivation for studying these stochastic sources is two-fold.  The
first reason is a rather hopeful one.  Because the gravitational force
is the weakest of the four known forces, the small-scale perturbations
of the gravitational field decouple from the evolution of the rest of
the universe at very early times.  Currently, our most detailed view of
the early universe comes from the microwave background radiation, which
decoupled from matter about $10^5$ years after the big bang, and gives
us an accurate picture of the universe at this rather early time.  Some
rather simple estimates (which I'll elaborate later) show that if the
current crop of gravity-wave detectors do detect a background of
cosmological origin, then it will carry with it a picture of the
universe as it was about $10^{-22}$ seconds after the big bang.  This
would represent a tremendous step forward in our knowledge, and is the
main reason for my interest.

The second reason for my interest in stochastic sources is rather more
practical.  One often hears it said that searching for signals in the
output of a gravitational wave antenna is like searching for a needle
in a haystack.  And indeed, for the first three types of source listed
above, this is true.  One has to search through tons of rock (the data
stream) in order to find the one precious gem (say, a binary chirp).
Most of the rock is barren, and the challenge is to isolate the one
tiny volume containing the material of interest.  As I have already
discussed, for gravitational wave sources, the amount of computational
power required for this careful search can be very large.  However the
situation is rather different when one is searching for a stochastic
background.  The analogy in this case is mining aluminum, where on
average every ton of ore contains a certain number of kilograms of
aluminum.  The situation is analogous for stochastic gravity-wave
sources.  As you will see, whenever two detectors are operating
simultaneously, even if only for a few seconds, we get a little bit
more data and information about the stochastic background.  And as you
will also see, it is easy to analyze this data.  The ``signal" in this
case is a very low bandwidth one, so the essential part of the data
analysis for a stochastic background can be done on a garden-variety
personal computer.  For example, in the case of the LIGO detectors, the
part of the signal carrying a significant amount of information lies
below a few hundred Hz (see Section~\ref{s:optimal} and
Fig.~\ref{f:filters} for details).   Thus the rate at which information
needs to be processed is only a few hundred data points/second; a very
manageable rate.

These lectures are organized as follows.  In Section~\ref{s:two}, I
discuss some of the general properties that a stochastic background of
gravitational radiation might have.  I show how such radiation is
characterized by a spectral function, discuss some of its statistical
properties, and show during which cosmological epoch the radiation
which falls into the bandwidth of the ground- and space-based detectors
was produced.  In Section~\ref{s:three} I show how one can combine data
from two or more gravity-wave detectors to either put limits on the
amplitude of a stochastic background, or to actually detect it.  I do
this in several steps, first giving a crude argument which demonstrates
the main detection strategy, then discussing the reduction in
sensitivity which comes about from the siting of the detectors and
their relative orientations on the earth.  This is followed by a
rigorous derivation of the optimal signal processing strategy, and a
calculation of the expected signal-to-noise ratio and the minimum
detectable energy-density in a stochastic background.  In
Section~\ref{s:what} I discuss the observational facts: what we
actually know about the stochastic background.  There are strong limits
on the spectrum of a stochastic background coming at very long
wavelengths from observations of the isotropy of the Cosmic Microwave
Background Radiation, and limits at higher frequencies from
observations of timing residuals in millisecond pulsars.  Finally,
there is a limit on the integrated spectrum arising from the standard
model of big bang nucleosynthesis.  The last part of these lectures is
far more speculative.  In Section~\ref{s:sources} I discuss some of the
potential ways in which a stochastic background might arise from
processes that take place early in the history of the universe.  I
discuss in some detail three particular models (inflation, cosmic
strings, and bubble formation in a first-order phase transition) each
of which gives rise to a different spectrum of radiation, and examine
in some detail the mechanisms at work in each case.
In each case I indicate if and
when the experiments being built might detect these potential sources.
This is followed by a short conclusion, and an appendix containing a
few calculational details.

\section{The stochastic background: spectrum \& properties}
\setcounter{equation}0
\label{s:two}
\subsection{Digression - the cosmic microwave background radiation}
\label{s:ssdig} 
In order to discuss the spectrum of gravitational background radiation,
we need to introduce notation.  I will do this with an analogy and an
example.

Let's begin by considering the electromagnetic background radiation,
conventionally referred to as the Cosmic Microwave Background Radiation
(CMBR) \cite{kolbturner}.  This radiation was originally produced when
the universe had a temperature of about $3000 \>{\rm K}$, at a redshift
of approximately $Z=1100$.  Today, this electromagnetic radiation has a
Planck blackbody spectrum with a temperature of about $T=2.73 \> {\rm
K}$.  Each cubic centimeter around us contains $\approx 400$ CMBR
photons of energy $\approx 10^{-15} \>{\rm erg}$.  The total energy
density in this radiation field is $\rho_{\rm em} = 4.2 \times 10^{-13}
\> {\rm ergs/cm}^3$, and the characteristic frequency of the photons
that make it up is (a few times) $f_0=k T/h=5.7 \times 10^{10} \> {\rm Hz}$.  Here $h
= 6.6 \times 10^{-27} \> \rm erg-sec$ is Planck's constant,
and $k=1.4 \times 10^{-16} \> \rm erg/Kelvin$ is Boltzmann's constant.

In order to characterize the spectral properties of this radiation,
we'll now consider how this energy is distributed in frequency.  In a
spatial volume $V$, the amount of energy $dE$ contained in this
radiation field between frequencies $f$ and $f+df$ can be expressed as
\be \label{e:diff}
dE = (2) hf  \left( {1 \over e^{hf/kT} - 1} \right) \left( {4 \pi V f^2
df \over c^3} \right),
\ee
where $c=3 \times 10^{10} \> \rm cm/sec$ is the speed of light.
The different factors appearing on the right hand side of this equation
are (1) the number of polarizations (2) the energy per quanta (3) the
number of quanta per mode, and (4) the number of modes in the frequency
interval.  Dividing both sides of this equation by the volume $V$, one
may write the energy density within the frequency range $df$ as
\be \label{e:edensity}
d \rho_{\rm em} \equiv {d E \over V} = {8 \pi h \over c^3} { f^3 df
\over e^{hf/k T} - 1} .
\ee
It will prove convenient to write the differential energy density
within a unit {\it logarithmic} frequency interval.  This is
\be \label{e:diffe}
{d \rho_{\rm em} \over d \ln f} = f {d \rho_{\rm em} \over d f} \approx
3.8 \times 10^{-14} {\rm ergs \over cm^3} \left( {f \over f_0 }
\right)^4 \left({ e-1 \over e^{f/f_0} - 1} \right).
\ee
This formula contains complete information about the spectral
distribution of energy in the CMBR, and if we were interested in
discussing electromagnetic backgrounds, we would probably stop here.
However there is a slightly different standard convention used to
describe gravitational wave backgrounds.

In describing gravitational wave stochastic backgrounds, it is
conventional to {\it compare} the energy density to the critical energy
density $\rho_{\rm critical}$ required (today) to close the universe.
This critical energy density is determined by the rate at which the
universe is expanding today.  Let us denote the Hubble expansion rate
today by
\be \label{e:hubble}
H_0 = h_{100} \> 100 \> {\rm  Km \over sec-Mpc} = 3.2 \times 10^{-18}
h_{100} {\rm 1 \over sec}  = 1.1 \times 10^{-28} c h_{100} {\rm 1 \over
cm }.
\ee
Because we don't know an accurate value for $H_0$ (a matter of
considerable controversy in the literature) we include a dimensionless
factor of $h_{100}$ which almost certainly lies within the range $1/2 <
h_{100} < 1$.  The critical energy-density required to just close the
universe is then given by
\be \label{e:crit}
\rho_{\rm critical} = { 3 c^2 H_0^2 \over 8 \pi G} \approx 1.6 \times
10^{-8} h_{100}^2 \rm \> ergs/cm^3.
\ee
This leads to our fundamental definition in this section, of a quantity
which we will be using for the remainder of these lectures.

\begin{figure}
\centerline{\epsfig{file=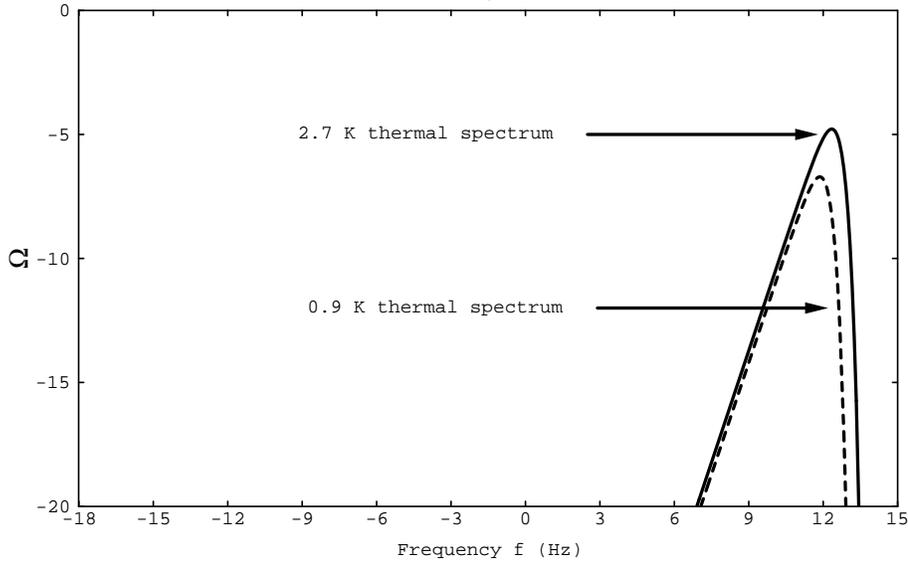,width=12cm,bbllx=72pt,bblly=240pt,
bburx=540pt,bbury=540pt}}
\caption{
\label{f:cmbr}
The solid curve is the fractional energy density $\Omega_{\rm em}(f)$
contained in the $2.73$ K electromagnetic background radiation (with
$h_{100}$ set to unity).  The dashed curve shows the corresponding
quantity for a $0.9$ K blackbody.  If the gravitational perturbations
had been in equilibrium with the matter fields, this is the expected
spectrum of the gravitational wave stochastic background.  Both axes
are $\rm log_{10}$.}
\end{figure}

In the remainder of these lectures, we will be discussing gravitational
wave, rather than electromagnetic backgrounds.  However, to complete
this section, we first define the analogous quantity for
electromagnetic backgrounds.  This quantity is a dimensionless function
of frequency
\be \label{e:defomega}
\Omega_{\rm em}(f) \equiv {1 \over \rho_{\rm critical}} {d \rho_{\rm
em} \over d \ln f}.
\ee
For the thermal spectrum of electromagnetic radiation in the CMBR, we have
\be \label{e:omegacmbr}
\Omega_{\rm em}(f) = 2.4 \times 10^{-6} h_{100}^{-2} \left( {f \over
f_0 } \right)^4 \left({ e-1 \over e^{f/f_0} - 1} \right).
\ee
A graph this function is shown in Fig.~\ref{f:cmbr}.  Also shown in
Fig.~\ref{f:cmbr} is the spectrum that the gravitational wave
stochastic background would have, if at early times in the history of
the universe the fluctuations in the gravitational field had been in
equilibrium with the other matter and radiation in the universe.  In
this case, the gravitational wave stochastic background would have a
thermal spectrum, with a temperature of about $0.9$ K
\cite{kolbturner}.  It is smaller than the temperature of the CMBR
because in a conventional hot big bang model, the gravitons would have
decoupled when the temperature of the universe dropped below the Planck
temperature, when the number of entropy degrees of freedom was $106.75$
in the standard GUT model.  Since the number of degrees of freedom
today is only 3.91, the graviton temperature is less than that of the
CMBR by the ratio $\left(106.75/3.91 \right)^{1/3}$.  (See discussion
in Kolb \& Turner (1990), pg 75, between (3.89) and (3.90)).  However it is
unlikely that this equilibrium could have been established;  the time
required to establish the equilibrium is longer than the characteristic
expansion time (the Hubble time) of the universe because the
gravitational interaction is so weak.  While it is therefore unlikely
that this $0.9$ K thermal spectrum is present, it is nevertheless a
useful benchmark for comparison.

A number of graphs similar to Fig.~\ref{f:cmbr} will appear in these
lectures, so a couple of comments are in order.  First, the reader will
notice that the horizontal axis encompasses an {\it enormous} range of
frequencies.   The lowest frequencies $f \approx H_0$ are those of
waves which only oscillate a single time in the entire history of the
universe, and whose period is a Hubble time $H_0^{-1}$!  The highest frequencies
shown are those of visible light.  Second,  these graphs make it easy
to see the total amount of energy contributed by the radiation to the
energy density of the universe.  From the vertical axis, one can
immediately see from the graph of $\Omega_{\rm em}(f)$ that the CMBR
contains, in the vicinity of $10^{11}$ Hz, about $10^{-5}$ of the
energy required to close the universe.

\subsection{
Notation - the spectral function $\Omega(f)$ for gravitational waves}
In order to characterize the spectrum of a stochastic gravitational
wave signal, we introduce a quantity for the graviton background which
is analogous to $\Omega_{\rm em}$.  The precise definition is
\be \label{e:defomegagw}
\Omega(f)=\Omega_{\rm gw}(f) \equiv {1 \over \rho_{\rm critical}} {d \rho_{\rm
gw} \over d \ln f}.
\ee
The subscripts ``gw" for ``gravitational wave"
are omitted when there is no danger of ambiguity.

There appears to be some confusion about $\Omega(f)$ in the
literature.  Some authors assume that $\Omega(f)$ is independent of
frequency $f$; this is true for some cosmological models, but not for
all of them.   The important thing is that {\it any} spectrum of
gravitational radiation can be described by an appropriate
$\Omega(f)$.  With the correct dependence on frequency $f$ it can
describe a flat spectrum, or a black-body spectrum, or any other
specific distribution of energy with frequency.

You will also notice that it follows directly from the definition that
the quantity $h_{100}^2 \Omega(f)$ is independent of the actual Hubble
expansion rate.  For this reason, we will often focus attention on that
quantity, rather than on $\Omega(f)$ alone.

It is sometimes convenient to discuss the spectrum of gravitational
waves and the sensitivity of detectors in terms of a characteristic
``chirp" amplitude.  This is the dimensionless gravitational-wave
strain $h = \Delta L/L$ that would be produced in the arms of a
detector, in a bandwidth equal to the observation frequency.  This is
related to $\Omega(f)$ by
\be \label{e:chirpamp}
h_{\rm c}(f) = 3 \times 10^{-20} h_{100} \sqrt{\Omega(f)} {{\rm 100
\> Hz} \over f}.
\ee
(Equation 65 of Thorne (1987)). 
Apart from an overall factor, this formula can be easily derived by
dropping the time derivatives from (\ref{e:rhoingw}).  For example if
$\Omega(f) = 10^{-8}$ over a bandwidth $50 \> {\rm Hz} < f < 150 \>
{\rm Hz}$ then the strain in an ideal detector is $h_{\rm c} \approx
10^{-24}$.

\subsection{
Assumptions about the stochastic gravitational-wave background}
Does $\Omega(f)$ contain all information about the stochastic
background?  The answer is ``yes" provided that we make enough
additional assumptions.

We will assume from here on that the stochastic gravity wave background
is isotropic, stationary, and Gaussian; under these conditions it is
completely specified by its spectrum $\Omega(f)$.  Each of these three
properties might or might not hold; before moving on, let's consider
each in turn.

It is now well established that the CMBR is highly isotropic
\cite{kolbturner}.  In fact this isotropy is surprising, because in the
standard model of cosmology, the angular size of the horizon at
$Z=1100$ is only about $2^\circ$.  Nevertheless, it is experimentally
well-established that the largest deviation from the isotropy of the
CMBR arises from our proper motion with respect to the rest frame of
the universe, at the level of 1 part in $10^3$.  The next largest
deviations from isotropy arise at the level of 1 part in $10^5$; these
fluctuations arise because of the non-uniform distribution of matter at
(and after) the surface of last scattering.

It is therefore not unreasonable to assume that the stochastic gravity
wave background is also isotropic. However this assumption may not be
true.  For example, suppose that the dominant source of stochastic
gravity wave background is a large number of unresolved white dwarf
binaries within our own galaxy.  Because our galaxy is bar or spiral
shaped (and not spherical) if we assume that the white dwarf binaries
are distributed in space in the same way as the matter in the galaxy,
then the stochastic background will have a distinctly anisotropic
distribution, and will form a ``band in the sky" distributed roughly in
the same way as the milky way. It is also possible for a stochastic
gravity wave background of cosmological origin to be quite
anisotropic.  Recent work \cite{allenottewill} has shown how data from
the LIGO detectors could be analyzed to search for such anisotropies.
Of course in this case, one is left with the problem of explaining why
the CMBR is isotropic, but the gravity wave background is not.  It is
possible to conceive of such mechanisms, but within the scope of these
lectures, we will not consider their effects.

The assumption that the stochastic background is stationary is almost
certainly justified.  Technically this means that the $n$-point
correlation functions of the gravitational wave fields depend only upon
the differences between the times, and not on the choice of the time
origin.  Because the age of the universe is 20 orders of magnitude
larger than the characteristic period of the waves that LIGO, VIRGO,
and the other facilities can detect, and 9 orders of magnitude larger
than the longest realistic observation times, it seems very unlikely
that the stochastic background has statistical properties that vary
over either of these time-scales.

The final assumption we make is that the fields are Gaussian, by which
we mean that the joint density function is a multivariate normal
distribution.  For many early-universe processes which give rise to a
stochastic background, this is a reasonable assumption, which can be
justified with the central limit theorem.   Let's sketch the argument
out.

\begin{figure}
\centerline{\epsfig{file=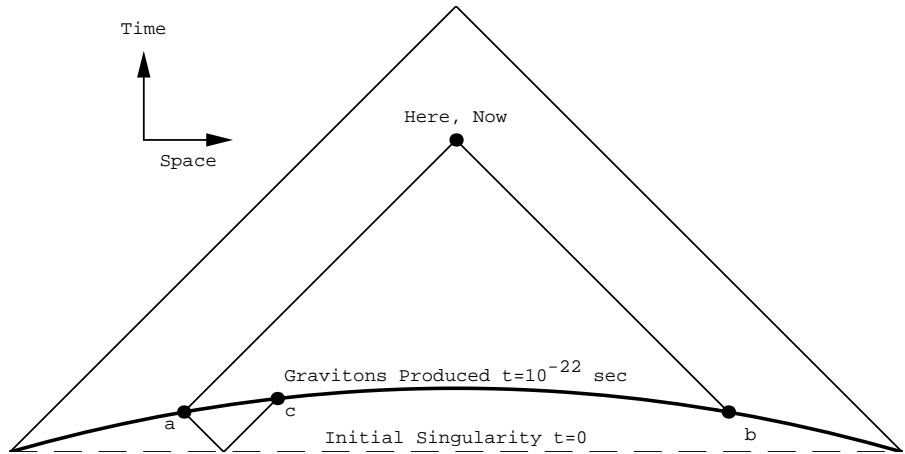,width=12cm,bbllx=2.5cm,bblly=9.8cm,
bburx=19.2cm,bbury=18.2cm}}
\caption{
\label{f:conf}
A conformal diagram showing a spatially-flat Friedman-Robertson-Walker
cosmological model.  The past light cone of a present-day observer
(Here, Now) intersects a surface at $t=10^{-22}$ seconds in a large
2-sphere, of which two points (a,b) are shown.  The future light cone
of the initial singularity intersects the same surface in a much
smaller 2-sphere, of which two points (a,c) are also shown.  The number
of independent, uncorrelated horizon volumes $N_{\rm horizon}$ which
contribute to the gravitational radiation arriving at a detector today
is given by the ratio of the areas of the larger 2-sphere to the
smaller one.}
\end{figure}

We will show in Section~\ref{s:whattime} that if the stochastic
background arises from processes that take place in the early universe,
the characteristic time (proper time after the big bang) at which the
gravitational radiation was emitted was about $t=10^{-22}$ sec.
Consider the conformal diagram \cite{hawkingellis}, shown in
Fig.~\ref{f:conf}.  As Fig.~\ref{f:conf} graphically illustrates, a
detector today located at the spacetime point labeled ``Here, Now"
observes radiation produced at $t=10^{-22}$ seconds after the big bang
by an extremely large number $N_{\rm horizon}$ of independent horizon
volumes.  We can estimate the number of these horizon-sized volumes on
the surface $t=10^{-22}$ seconds as follows.

Let's assume that the universe is $k=0$ (spatially flat) and
radiation dominated from the time that the gravitons were produced
($t_1=10^{-22}$ sec) until the present time ($t_0=10^{17}$ sec).  This
is a reasonable assumption for this type of calculation; although the
universe did recently become matter dominated (at a redshift $Z$ of a few
thousand) this has only a small effect on the final answer.   The
redshift $Z$ of the $t=t_1$ constant-time surface is then given by $1+Z
= (10^{17}/10^{-22})^{1/2} \approx 10^{20}$.  Let's work in a
``conformal time" coordinates where $ds^2 = a^2(\eta)(-d\eta^2 + d\vec
x^2)$.  For a radiation-dominated universe $a(\eta) = \eta$, and we can
take the present time to have $\eta=\eta_0$ and the time at which
gravitons were produced to have $\eta=\eta_1$.  We begin by considering
the intersection of the forward light cone of the initial singularity
with the surface $\eta = \eta_1$.  This intersection forms the small
two-sphere denoted by the points ``a" and ``c" in Fig.~\ref{f:conf};
this two-sphere has area $A_{\rm small}= 4 \pi \eta_1^2 a^2(\eta_1)$.
In similar fashion, the past light cone of the present-day detector
intersects the surface $\eta=\eta_1$ in the large two sphere denoted by
the points ``a" and ``b", which has area $A_{\rm big}=4 \pi
(\eta_0-\eta_1)^2 a^2(\eta_1)$.  The ratio of these two areas is the
number of horizon volumes visible to the detector over the entire sky;
it is
\be \label{e:nhorizon} 
N_{\rm horizon} = {A_{\rm big} \over A_{\rm small}} = {(\eta_0 -
\eta_1)^2 \over \eta_1^2 } = Z^2 \approx 10^{39}.
\ee
If we assume that the processes producing gravitational waves from each
of these separate horizon volumes act independently, then it follows
immediately from the central limit theorem that the amplitude of the
radiation arriving at the detector, which is the sum of the amplitudes
of the radiation produced by each of the separate volumes, is
Gaussian.

\subsection{
When is a stochastic background produced?}
\label{s:whattime} 
Suppose that the LIGO or VIRGO detectors do detect a stochastic
background of gravitational radiation.  From what epoch, in the
history of the universe, does this radiation date?

To give a general answer to this question, we need to make some
assumptions about the universe.  The most reasonable thing to do is to
adopt the ``standard model" of cosmology.  This cosmological model
consists of a spatially-flat ($k=0$ Friedman-Robertson-Walker)
cosmological model, in which the equation of state is dominated by
massless (or highly relativistic) matter for redshifts greater than
$Z_{\rm eq} \approx 6000$, and is dominated by massive pressureless
particles (dust!) for smaller redshifts.

The question posed above has two answers, one mundane, the other quite
exciting.  The mundane possibility is that the radiation was produced
at or near the present epoch (say, within a redshift $Z<4$) by many
unresolved separate sources, such as white dwarf binaries or
supernovae.  For example, in the case of the electromagnetic background
radiation, one finds that below about $300 \> \rm MHz$ the spectrum is
dominated by (recent) emission from our own galaxy \cite{weinberg}.
The more exciting possibility is that such gravitational radiation is a
cosmological fossil.  In the case of the electromagnetic spectrum, at
frequencies above 1 GHz, this is indeed the case.

Note that we must not discount the possibility that the stochastic
background in the LIGO frequency band was produced near the present
epoch.  Recent work \cite{blair} has shown that if the efficiency with
which supernovae convert their mass into gravitational radiation
exceeds $10^{-5}$ then the cosmological distribution of such supernovae
could be a dominant (and detectable) source of stochastic background in
the LIGO frequency band.
 
\begin{figure}
\centerline{\epsfig{file=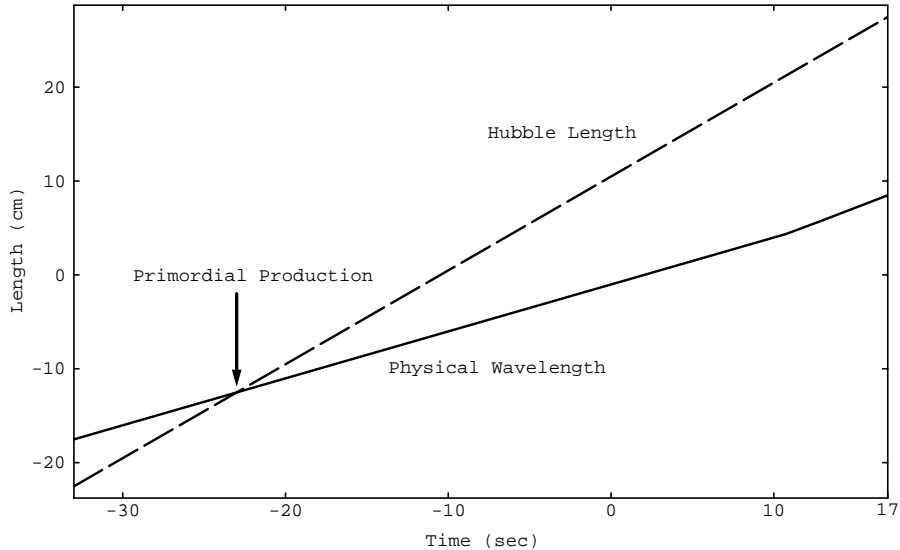,width=12cm,bbllx=2.5cm,bblly=8.7cm,
bburx=19.2cm,bbury=19.2cm}}
\caption{
\label{f:prod}
This graph shows the two lengths, as functions of cosmological time
(both axes are ${\rm log}_{10}$).  Today, we are at the far right of
the graph, $t \approx 10^{17}$ seconds after the big bang.  The solid
curve shows the physical wavelength of a wave that LIGO or VIRGO might
detect, at $\approx 100 \> \rm Hz$.  Today this wavelength is about $3
\times 10^8$ cm; in the past this wavelength shrinks because of
redshifting, in proportion to the cosmological scale factor, first as
$t^{2/3}$ when the universe is matter dominated, then as $t^{1/2}$ when
the universe is radiation dominated.  The dotted curve shows the
characteristic spatial size (and age) of the universe, the Hubble
length, which today is $\approx 10^{28}$ cm.  This function
grows $\propto t$.  The intersection of these two curves determines the
time at which the primordial production of gravitons in the LIGO/VIRGO
band took place.
}
\end{figure}

Let us assume that the LIGO or VIRGO detectors do detect a stochastic
background of gravitational radiation, and let us also assume that this
radiation has a cosmological (a long time ago) rather than a local (recently)
origin.  The time at
which this radiation was produced can be determined by simple physical
argument.  The production of gravitational
radiation is no different than the production of any other type of
radiation.  If we want to produce electromagnetic radiation at $1$ KHz,
we need to take electric charges and vibrate them at $1$ KHz.  The same
holds for gravitational radiation; waves of a certain frequency are
produced when the characteristic time for the matter and energy in the
universe to shift about is comparable to the period of the waves.   The graph in Fig.~\ref{f:prod} shows two length scales, as
functions of cosmic time.  As
can be seen from the diagram, at about
$10^{-22}$ seconds after the big bang the
characteristic expansion time (the Hubble time) was about comparable to
the period of a wave, which when redshifted to the present epoch,
falls within the bandwidth of LIGO/VIRGO sensitivity.

The most exciting prospect in the study of the stochastic background is
the possibility of detecting radiation produced at this very early
epoch of $t \approx 10^{-22}$ sec.  To give a sense for the scales, a
100 Hz graviton detected by the LIGO experiment was produced at that
early epoch with an energy of order $10$ MeV.  The mean temperature in
the universe at that time was of order $10^7$ GeV.  For this reason,
detecting a background of cosmological origin would give us a glimpse
of the universe at {\it much} earlier times than we can obtain in other
ways.

Comparable calculations can be done for the proposed Laser
Interferometer Space Antenna (LISA) experiment \cite{lisa}, which is
designed to detect waves with frequencies in the range $\approx 10^{-4}
\> \rm Hz$ to $\approx 10^{-1} \> \rm Hz$.  In this case, one finds
that waves of cosmological origin with present-day frequency $\approx
10^{-2} \> \rm Hz$ were produced at time of order $10^{-14} \> \rm sec$
after the big bang; at the time of their production the energy of the
gravitons was $\approx 0.1 \> \rm eV$.  The overall temperature of the
universe at this time was $\approx 10^3 \> \rm GeV$ - a particularly
interesting epoch, because the electro-weak phase transition ocurred
around that energy scale.

\section{
\label{s:three} 
 Experimental detection methods and their sensitivity}
\subsection{How to detect a stochastic background (crude version)}
\setcounter{equation}0
\label{s:how} 
In these lectures, I'd like to answer the question ``how does one
actually use the output from the gravitational wave detectors to find
(or place upper limits on) a stochastic gravity wave background."
Later, in Section~\ref{s:optimal} we will re-visit this question and
give a very precise (in fact, optimal) answer to this question.
However my goal here is to give a rough understanding of the method and
its limitations.

Now in general, there are two possibilities.  First, when a detector is
operating, its intrinsic noise level might be lower than the random
stochastic signal arising from the gravitational wave background.  In
this case, no clever signal detection strategies are required.  One
simply ``turns on" the detector, studies its output for a few seconds,
and concludes that a stochastic background is present.  In a crude
sense, this is how Penzias and Wilson discovered the electromagnetic
CMBR; the signal it produced was greater than the intrinsic noise of
their detector.  Of course it was not quite so easy - Penzias and
Wilson had to spend an entire summer studying the intrinsic noise in
their radio telescope before they could be sure that the excess noise
they were observing was of cosmological origin!

Let us assume that the signal levels produced by the stochastic
background of gravitational radiation are not this large, so that the
output of a detector is dominated by the noise produced within the
detector rather than by the signal due to the stochastic background
itself.  In this case, as was first shown by Michelson (1987) one
can still detect the signal due to a stochastic background, by
correlating the outputs of two different detectors.  The principal
requirement that must be met by the two detectors is that they must
have {\it independent noise}.  Later in these lectures we will be more
specific about what is required.

\begin{figure}
\centerline{\epsfig{file=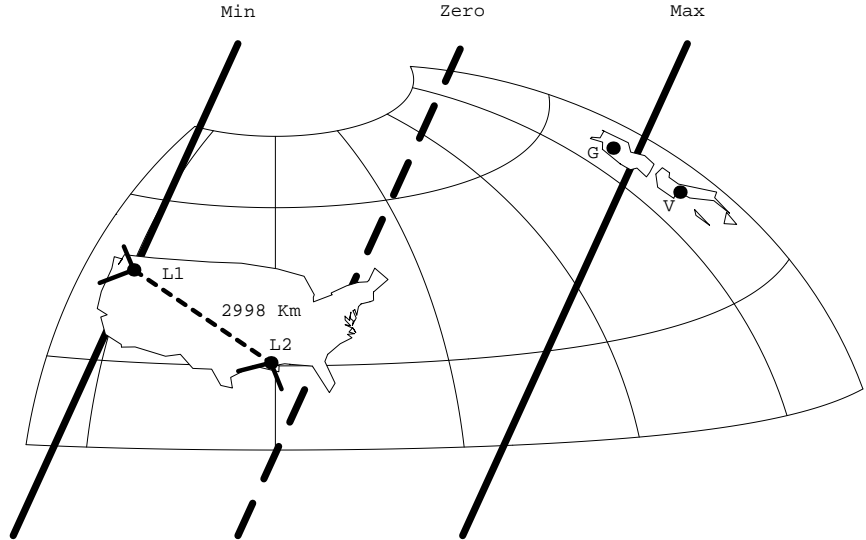,width=12cm,bbllx=2.5cm,bblly=8.7cm,
bburx=19.2cm,bbury=19.2cm}}
\caption{ \label{f:earth} Shown is a the surface of the earth $15^\circ
< \rm latitude  < 75^\circ$, and $-130^\circ < \rm longitude <
20^\circ$, including the LIGO detectors in Hanford, WA (L1) and
Livingston, LA (L2) and the locations of the VIRGO detector (V) in
Pisa, Italy and the GEO-600 (G) detector in Hannover, Germany.  The
perpendicular arms of the LIGO detectors are also illustrated (though
not to scale).  A plane gravitational wave passing by the earth is
indicated by successive minima (troughs), zeros, and maxima (peaks) of
the wave.  As this wave passes by the pair of LIGO detectors, it
excites the detectors in coincidence at the moment shown, because both
detectors are driven negative by the wave.  During the time when the
zero is between L1 and L2, the two detectors respond in
anti-coincidence.  Provided that the wavelength $\lambda$ of the wave
is larger than about twice the separation distance (2998 Km) between
the detectors, on the average they are driven in coincidence more of
the time than in anti-coincidence.}
\end{figure}

In the case where we have two detectors, one can isolate the signal due
to the stochastic background by correlating their outputs.  As can be
seen from Fig.~\ref{f:earth}, a gravitational plane wave passing by the
earth excites a pair of detectors in coincidence when the positive or
negative amplitude part of the wave is passing by both detectors at the
same time (in the rest frame of the detectors).  It excites the
detectors in anti-coincidence during moments when when the positive
amplitude part of the wave is passing by one detector, and the negative
amplitude part of the wave is passing by the other detector, or vice
versa.  Provided that the wavelength of the wave is much larger than
twice the distance between the two detectors, on the average the
detectors will be driven in coincidence.  In the case of the LIGO
detector pair, below a frequency of $64$ Hz, the Hanford and Livingston
detectors will be driven (on the average) in coincidence by an
unpolarized isotropic stochastic background.  This can be seen by
inspecting Fig.~\ref{f:overlap}; the first zero of this function lies
at $f=64$ Hz.   We note that this result is slightly higher than the
frequency (50 Hz) associated with a wavelength of twice the detector
separation (6000 Km) because the overlap reduction function is a sum of
Bessel functions (\ref{e:bessels}) which does not vanish at 50 Hz.

The basic idea (the optimal implementation is elaborated in
Section~\ref{s:optimal}) is to multiply together the outputs of the two
independent detectors, and integrate the result.  For the purpose of
this section, imagine that two detectors are located at the same point,
and have identical orientations.  In our case, for frequencies well
below $64$ Hz, the LIGO detectors may be thought of as ``being at the
same point".  In similar fashion, although the LIGO detectors are not
identically oriented (they are $27.2^\circ$ apart as seen from the
center of the earth, and thus can not lie in the same plane) their arms
are as close to parallel as is possible given their separation.

In this simplified case the output of the first detector is
\be \label{e:sigdef1}
s_1(t) = h_1(t) + n_1(t),
\ee
where $h_1$ is the strain due to the stochastic background and $n_1$ is
the intrinsic noise of the first detector.  In similar fashion, the
output of the second detector is
\be \label{e:sigdef2}
s_2(t) = h_2(t) + n_2(t).
\ee
Since we are assuming that the two detectors have identical locations
and orientations then the gravity-wave strains are identical:
$h_1(t)=h_2(t)$.  We can form a ``signal" by multiplying together the
outputs of the two detectors and integrating:
\be 
S = <s_1,s_2> \equiv \int_{-T/2}^{T/2} s_1(t) s_2(t) dt.
\ee
Here $T$ is the integration time (realistically, a few months).  There
are two different cases to consider.  As we explained earlier, the case
where the detector noise $n_i$ is small compared to the signal $s_i$
not very interesting, as one can find the background easily in that
case.  So we'll assume that the signal is small compared to the
detector noise, in which case we can write
\bea \label{e:neglect}
S &= &<h_1, h_2> + <n_1,h_2> + <h_1,n_2> + <n_1,n_2> \\
&\approx &<h_1, h_2> + <n_1,n_2>.
\eea
where we have dropped terms like $<n_1,h_2>$ that are smaller than
$<n_1,n_2>$ but statistically identical, because $n_1$ and $h_2$ are
uncorrelated in the same way as $n_1$ and $n_2$.  (To simplify, we
assume that there are no correlated sources of noise such as seismic
activity).

The two terms appearing in the signal $S$ grow in different ways as the
observation time $T$ increases.  The first term $<h_1,h_2>$ increases
linearly with time $T$, because $h_1=h_2$.  The second term $<n_1,n_2>$
can be thought of as a random walk on a one-dimensional line, and so on
the average it grows $\propto \sqrt{T}$.  We will show in
Section~\ref{s:optimal} that
\bea
<h_1,h_2>& \propto &| \tilde h(f) |^2 \Delta f T \propto \Omega(f) \Delta f T\\
 <n_1,n_2>& \propto&  |\tilde n(f)|^2 \sqrt{\Delta f T}
\eea
where $f$ is the central frequency at which the detector is sensitive,
$\Delta f$ is the effective bandwidth, and $\tilde Q$ is the Fourier
transform of $Q$.  Because the ``signal" $<h_1,h_2>$ grows like $T$,
and the noise $<n_1,n_2>$ only grows as $T^{1/2}$, with enough
observation time one can in principle detect a gravitational wave
stochastic background buried in {\it} any level of detector noise.
Setting the ``signal" term greater than the noise term gives a minimum
detectable level of $\Omega$:
\be
\Omega_{\rm minimum \> detectable} \propto {|\tilde n(f)|^2 \over \sqrt{\Delta f T}}.
\ee
Thus, even if the detector noise is much larger than the expected
stochastic gravity wave strain, by making a long enough observation,
one can in principle observe even very low levels of stochastic
background.

Let me stress one more time that this crude analysis is intended
primarily to present the concepts and basic ideas, and will be treated
rigorously in Section~\ref{s:optimal}.

\subsection{The overlap reduction function $\gamma(f)$}
\label{s:overlap} 
To provide a rigorous treatment of the signal analysis, we must take
into account the reduction in sensitivity that results from two effects
(1) the non-parallel alignment of the arms, and (2) the separation time
delay between the two detectors.  These two effects reduce the
sensitivity of a stochastic background search;  they mean that $h_1$
and $h_2$ are no longer equal; the overlap between the gravity wave
strains in the two detectors is only partial.

To quantify these effects we introduce the {\it overlap reduction
function} $\gamma(f)$ first calculated in closed form by Flanagan
(1993).  This is a dimensionless function of frequency $f$, which
depends entirely on the relative positions and orientations of a pair
of detectors.  The overlap reduction function $\gamma(f)$ equals unity
for co-aligned and coincident detectors, and decreases below unity when
the detectors are rotated out of co-alignment (so they are sensitive to
different polarizations) or shifted apart (so that there is a
phase-shift between the signals in the two detectors).  Later in these
lectures (in Section ~\ref{s:optimal}) we will see how this function
comes up naturally in the determination of signal-to-noise ratios.

The overlap reduction function is defined by \cite{flan}
\be
\label{e:overlap}
\gamma(f) \equiv {5 \over 8 \pi} \int_{S^2} d \hat \Omega \>
e^{2 \pi i f \hat \Omega \cdot \Delta \vec x/c }
\left( F_1^+ F_2^+ + F_1^\times F_2^\times \right).
\ee
Here $\hat \Omega$ is a unit-length vector on the two-sphere, $\Delta
\vec x$ is the separation between the two detector sites, and
$F_i^{+,\times}$ is the response of detector $i$ to the $+$ or $\times$
polarization.  For the first detector $(i=1)$ one has
\be \label{e:detectresponse}
F_1^{+,\times}
 = {1 \over 2} \left( \hat X_1^a \hat X_1^b - \hat Y_1^a \hat Y_1^b \right)
e_{ab}^{+,\times}(\hat \Omega),
\ee
where the directions of a single detector's arms are defined by $\hat X_1^a$
and $\hat Y_1^a$, and $e_{ab}^{+,\times}(\hat \Omega)$ are the spin-two
polarization tensors for the ``plus" and ``cross" polarizations
respectively.  These quantities are defined in equations
(\ref{e:polar},\ref{e:polar2}) of Appendix~\ref{s:appendix1}.

It is not hard to see how the overlap reduction function $\gamma(f)$
arises.  As we show in detail in Appendix~\ref{s:appendix1} $\gamma(f)$
is proportional to the averaged product of the strains at two different
detectors, when those detectors are driven by an isotropic unpolarized
stochastic background of gravitational radiation at frequency $f$.
This radiation is arriving at the detectors uniformly from all points
on the celestial sphere, hence the integral over the direction $\hat
\Omega$ of the wave-vector of the arriving radiation.

Let us look at each term in (\ref{e:overlap}).  (1) The overall
normalization factor is chosen so that if $| \Delta \vec x|=0$ (or
equivalently, $f=0$) and the two detectors are coincident, and the two
detectors have parallel pairs of arms, then $\gamma \rightarrow 1$.
(2) The exponential phase factor is the phase shift arising from the
time delay between the two detectors for radiation arriving along the
direction $\hat \Omega$.  (3) The quantity $F_1^+$ is the response of
the first detector to the ``$+$" polarization wave, and similarly for
detector 2 and the ``cross" polarization.

In Section~\ref{s:optimal} this overlap reduction function will appear
after a rigorous calculation; these remarks are meant primarily for
motivation.

\begin{figure}
\centerline{\epsfig{file=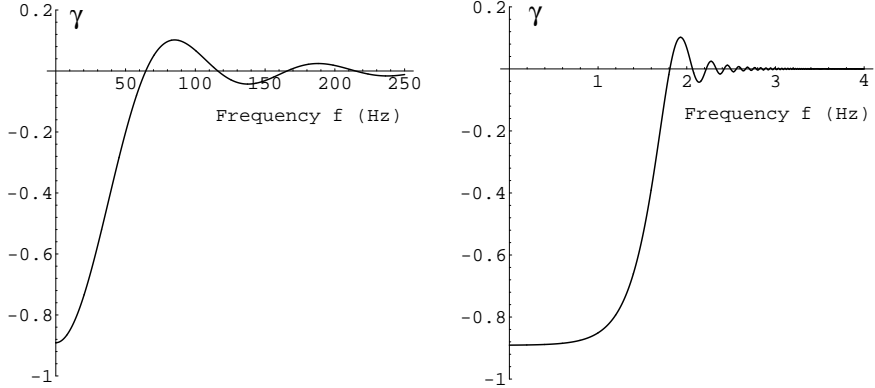,width=12cm,bbllx=2.8cm,bblly=10.3cm,
bburx=19.2cm,bbury=17.5cm}}
\caption{
\label{f:overlap}
The overlap reduction function $\gamma(f)$ for the two LIGO detector
sites.  (The horizontal axis of the left-hand graph is linear, while
that of the right-hand graph is $\rm log_{10}$.)  The overlap reduction
function shows how the correlation of the detector pair to an
unpolarized stochastic background falls off with frequency.  The
overlap reduction function has its first zero at 64 Hz, as explained
earlier.  It falls off rapidly at higher frequencies.}
\end{figure}

A graph of $\gamma(f)$ is shown in Fig.~\ref{f:overlap} for the two
LIGO sites in Hanford, WA and Livingston LA.  The overlap reduction
function is {\it negative} as $f \rightarrow 0$.  This is because,
comparing the two LIGO sites, the arm orientations are not parallel but
are rotated relative to one another by $90^\circ$.  This can be seen
immediately in Fig.~\ref{f:earth}.  If the Livingston detector arms
(denoted L2) were rotated by $90^\circ$ in the clockwise direction,
this would change the overall sign of $\gamma(f)$, but otherwise leave
its dependence on $f$ unchanged.  The magnitude of $\gamma(0)$ is not
unity, because the plane of the Hanford detector and that of the
Livingston detector are different.  Thus the arms of the two detectors
are not exactly parallel, and hence $|\gamma(0)|=0.89$ rather than
$1$.

The precise form of the overlap reduction function is not important
here.  However it can be calculated in closed form \cite{flan}, and may
be expressed as the sum of three Bessel functions:
\be \label{e:bessels}
\gamma(f) = A j_0(\alpha) + B {j_1(\alpha) \over \alpha} 
+ C {j_2(\alpha) \over \alpha^2}
\ee
where $\alpha=2 \pi f |\Delta \vec x|/c$ depends upon the detector
separation and the frequency, and $A,B,C$ are constants which only
depend upon the locations of the detectors and the orientations of
their arms \cite{flan}.
For example, for the two LIGO detectors, one finds $A=-0.124842$, $B= -
2.90014 $, and $C=3.00837$.

\subsection{Optimal filtering}
\label{s:optimal} 
I now return to the subject that we addressed in Section~\ref{s:how}:
how do we detect a stochastic background?  In this section, we work out
the precise answer to this question.  This type of analysis was first
carried out by Michelson (1987) then improved by Christensen
(1992) and further refined by Flanagan (1993).  The analysis
given here appears in Bayesian form in the work of
Flanagan (1993); the treatment given here is the somewhat simpler
``frequentist" version.  The two different approaches yield exactly the
same final results, because the signals are Gaussian.

Let us write the most general possible form of the correlation
``signal" between two detectors in the form
\be \label{e:sig1}
S \equiv \int_{-T/2}^{T/2} dt \int_{-T/2}^{T/2} dt' s_1(t) s_2(t') Q(t-t').
\ee
The strains $s_1$ and $s_2$ are the signal streams from (say) the
Hanford and Livingston LIGO detectors, defined by
(\ref{e:sigdef1}-\ref{e:sigdef2}), and $Q(t-t')$ is a real filter
function.  Because we are assuming that the statistical properties of
our signals and noise are stationary, the ``best choice" of filter
function $Q$ can only depend upon the difference $\Delta t = t - t'$ of
the two times.  Our goal, in this section, is to find the {\it optimal
choice} of filter function $Q(\Delta t)$ in a rigorous way.

The optimal choice of filter function $Q(\Delta t)$ depends upon the
locations and orientations of the detectors, as well as on the spectrum of
the stochastic background, and the noise characteristics of the
detectors.  If the two sites are very close together (compared to the
characteristic wavelength at which the detectors are sensitive), have
identical orientations, and the detectors have a narrow bandwidth, then
the optimal choice of filter is a Dirac delta function $Q(t-t') =
\delta(t-t')$.  In this case, we just multiply the signals and
integrate, precisely as discussed earlier in Section~\ref{s:how}.  Even
if {\it none} of these assumptions holds, you will see that the optimal
choice of filter function $Q$ falls rapidly to zero for time delays
$\Delta t$ whose magnitude is large compared to the light travel time
$d/c = 10^{-2}$ seconds between the two sites.  We are thus justified
in changing the limits above, to obtain
\be \label{e:sig1p}
S = \int_{-T/2}^{T/2} dt \int_{-\infty}^{\infty} dt' s_1(t) s_2(t') Q(t-t').
\ee
Since the observation time $T$ will be at least several months, one can
see immediately that the signal $S$ gets most of its support in the
integral (\ref{e:sig1}) from the region near the diagonal line $t=t'$
in the region $[-T/2,T/2] \times [-T/2,T/2]$.  Hence the value of $S$
grows linearly with integration time $T$, in perfect agreement with the
earlier ``intuitive" explanation in Section~\ref{s:how}.

The reader should note that with our definition, the ``signal" $S$ is a
quantity {\it quadratic} in the strain $h$; this is not the same as the
convention frequently followed in analyzing signals from chirp or
periodic sources, where the ``signal" is generally taken to be a
quantity {\it linear} in the strain $h$.  For this reason, the
expressions that we will derive for quantities such as the signal to
noise ratio are in some ways analogous to the {\it squares} of the
corresponding quantities for chirp or periodic signals.

We can re-write (\ref{e:sig1p}) in the frequency domain.  Our convention
for Fourier transforms is
\be
\tilde g(f) = \int_{-\infty}^{\infty} dt \> e^{-2 \pi i f t} g(t).
\ee
Here, and elsewhere in these lectures, ${}^*$ denotes complex
conjugation, and within this section one may assume that $s_i(t)$
vanishes at some very early time (say, before the detector was built)
and at some very late time (say, after the detector is decommissioned)
so that the Fourier transforms are well-defined.  Because we have
assumed that the filter function is real, $\tilde Q(-f) = \tilde
Q^*(f)$.  Identical reality conditions hold for the signals $\tilde
s_i$, the strains $\tilde h_i$, and the detector noise $\tilde n_i$.
Using the Fourier transform, we may write the signal (\ref{e:sig1p}) as
\be \label{e:sig2}
S = \int_{-\infty}^{\infty} df \> \int_{-\infty}^{\infty} df' \>
\delta_T(f-f') \tilde s_1^*(f) \tilde s_2(f') \tilde Q(f').
\ee
The function $\delta_T$ which appears on the r.h.s. of
(\ref{e:sig2}) is a ``finite time" approximation to the Dirac delta
function, defined by
\be 
\delta_T(f) \equiv \int_{-T/2}^{T/2} dt \> e^{ -2 \pi i f t}
= {\sin(\pi f T) \over \pi f},
\ee
which reduces to the Dirac $\delta$-function $\delta(f)$ in the limit
$T\to\infty$.  Note however that for finite observation times $T$ one
has $\delta_T(0) = T$.  We now consider the stochastic properties of the
signal $S$.  In order to find the optimal choice of the filter
function, some additional assumptions are needed.

As discussed in detail earlier, we will assume that the stochastic
background is isotropic, unpolarized, and Gaussian.  Under these
assumptions, the expectation value of the Fourier amplitudes of the
strain signal and the strain noise are derived in Appendix~{\ref{s:appendix1}}:
\bea \label{e:expected}
\langle \tilde h_1^*(f) \tilde h_2(f') \rangle &=& \delta(f-f') {3
H_0^2 \over 20 \pi^2} |f|^{-3} \Omega_{\rm gw} (|f|) \gamma(|f|), \> \>
{\rm and}\\
\label{e:expected2}
\langle \tilde n_i^*(f) \tilde n_j(f') \rangle &=& {1 \over 2} \delta(f-f')
\delta_{ij} P_i(|f|).
\eea
Here the indices $i,j$ label the sites, so for example $i=1$ is
Hanford Washington and $i=2$ is Livingston Louisiana. The power spectral noise density
function $P_i(f)$ which is defined by (\ref{e:expected2}) is a real
non-negative function, defined with a factor of $1/2$ to agree with the
standard definition used by instrument-builders, so that the total
noise power is the integral of $P(f)$ over all frequencies from $0$ to
$\infty$ (rather than starting at $-\infty$).  You will notice that we
have assumed that the detector noise at the two different sites is
completely uncorrelated.  This assumption significantly simplifies the
analysis, but is not entirely realistic (Christensen 1992; Flanagan 1993).

\begin{figure}
\centerline{\epsfig{file=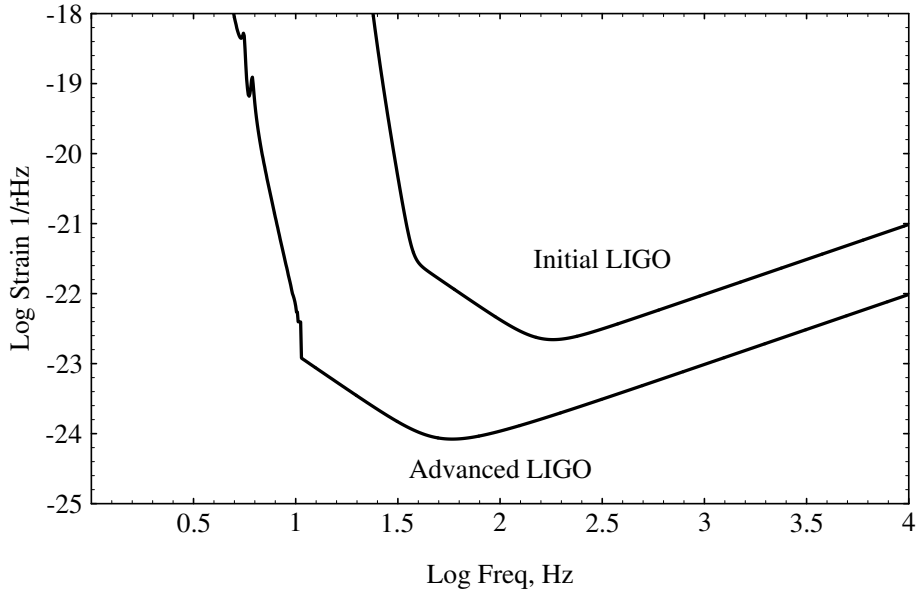, width=12cm, bbllx=74pt, bblly=248pt,
bburx=537pt, bbury=565pt}}
\caption{
\label{f:noise}
The predicted noise power spectra of the initial and advanced LIGO
detectors.   The horizontal axis is $\rm log_{10}$ of frequency $f$, in
Hz.  The vertical axis shows $\rm log_{10}  (P(f)/sec)^{1/2}$, or
strain per root Hz.  These noise power spectra are the published design
goals.  The bumps appearing in the low-frequency part of the advanced
LIGO noise curve are obtained by folding measured seismic noise data
with the predicted transfer function of the seismic isolation (stack)
system.}
\end{figure}

The predicted detector noise for the initial and advanced LIGO is shown
in Fig.~\ref{f:noise}.  This figure is taken from the LIGO design goals
\cite{science92}.  Because the 4 kilometer detectors are identical,
they have the same power spectral noise densities,
$P(f)=P_1(f)=P_2(f)$.   There are at least three different sources of
detector noise which dominate at different frequencies.  At the low
frequency end, the dominant source of noise is seismic noise.  In the
advanced version of LIGO, this noise will be reduced by improving the
mechanical filters that isolate the detector from the ground.   In the
mid-band region, the noise is primarily due to thermal motion of the
masses and mirrors, which will be reduced by using improved materials
and design.   In the high-frequency part of the spectrum, the noise is
dominated by photon shot noise, which can be reduced by increasing the
laser power and improving the optical components.

Although in the case of the LIGO detectors, the noise power spectra of
the detectors are expected to be the same, in general, we are
interested in correlating the outputs of detectors with different
designs (say the LIGO and VIRGO detectors).  So we will keep our
treatment fairly general, and treat $P_1(f)$ and $P_2(f)$ as distinct
functions.

In order to find the optimal choice of filter function $Q$, we need to
choose the quantity to maximize.  The natural choice here is to
maximize the ratio of signal-to-noise.  The expected value of the
signal is obtained by taking the expectation value of (\ref{e:sig2}),
making use of (\ref{e:expected}). 
Because the noise in each detector is uncorrelated with the other detector,
and is uncorrelated with the gravity-wave strain $h$, one obtains
\bea
\langle S \rangle &=& \int_{-\infty}^{\infty} df \> 
\int_{-\infty}^{\infty} df' \> \delta_T(f-f') \langle \tilde
h_1^*(f) \tilde h_2(f') \rangle \tilde Q(f') \\
\label{e:sig3}
&= &{3 H_0^2 \over 20 \pi^2} T
\int_{-\infty}^{\infty} df \> \gamma(|f|) |f|^{-3} \Omega_{\rm gw} (|f|)
\tilde Q(f). \label{eq30}
\eea
The factor of $T$ on the r.h.s. arises from evaluating $\delta_T(0)$.
To determine the amount of noise $N$ in the detector, we consider the
variation of the signal $S$ away from its mean value:
\be
N \equiv  S- \langle S \rangle.
\ee
The quantities on the r.h.s. are easily determined, if we assume as in
(\ref{e:neglect}) that the noise $n_i(t)$ each detector is much larger
in magnitude than the gravitational strain $h_i(t)$.  In this case, one
may simply replace the detector output signal (gravity strain + noise)
by the detector noise alone in (\ref{e:sig2}), obtaining
\be \label{e:noise}
N \approx \int_{-\infty}^{\infty} df \> \int_{-\infty}^{\infty} df'
  \> \delta_T(f-f') \tilde n_1^*(f) \tilde n_2(f') \tilde Q(f').
\ee
This approximation is consistent with the definition of $N$; a brief
calculation shows that the expected value of $N$ vanishes, because the
noise in the two detectors is assumed to be uncorrelated,
\be
\langle N \rangle = 0,
\ee
a result which also follows directly from the definition of $N$.
However the r.m.s. value of the expected noise is not zero, because if
we take the expectation value of the square of (\ref{e:noise}) and use
(\ref{e:expected2}) one obtains
\bea
\langle N^2 \rangle &= & \langle S^2 \rangle - \langle S \rangle^2 \\ &
= &{1 \over 4} \int_{-\infty}^\infty df \> \int_{-\infty}^\infty df' \>
\delta_T^2(f-f') P_1(|f|) P_2(|f'|) \left| \tilde Q (f') \right|^2.
\eea
For long observation times (say $T > 1 \> \rm sec$) the finite-time
delta function $\delta_T(f-f')$ is sharply peaked over a region in
$f-f'$ whose size $\approx 1/T$ is very small compared to the scale on
which the functions $P_1, P_2, \tilde Q$ are changing.  In this case we
are justified in replacing one of the finite-time delta functions above
by a Dirac delta function, obtaining
\be 
\langle N^2 \rangle = 
{T \over 4}
\int_{-\infty}^\infty df \> P_1(|f|) P_2(|f|) \left| \tilde Q (f)
\right|^2.
\ee
Notice that the expected r.m.s. noise $\langle N^2 \rangle^{1 \over 2}$
grows as $\sqrt{T}$, in complete agreement with the simple intuitive
explanation given earlier in Section~\ref{s:how}.

The problem now is to find
the choice of filter function $\tilde Q$ which maximizes the signal to
noise ratio
\be
\left({S \over N}\right) \equiv {\langle S \rangle \over \langle N^2
\rangle^{1 \over 2}}.
\ee
This turns out to be remarkably straightforward, if we introduce an
inner product between two functions of frequency, $A(f)$ and $B(f)$.
This inner product takes as inputs two arbitrary complex functions
$A(f)$ and $B(f)$; the output $(A,B)$ is a single complex number.  We
define this inner product by
\be
(A,B) \equiv \int_{-\infty}^{\infty} df \> A^*(f) B(f) P_1(|f|) P_2(|f|).
\ee
It is straightforward to see that this inner product is positive
definite; since $0 < P_i(f)$ the norm of a function $(A,A)$ vanishes if
and only if $A(f)$ vanishes.  In terms of this inner product, we can
express the expected signal and noise as
\bea
\langle S \rangle & = &( \tilde Q, {\gamma(|f|) \Omega_{\rm gw}(|f|) \over |f|^3
P_1(|f|) P_2(|f|)} ) \> {3 H_0^2 \over 20 \pi^2} T , \> \> {\rm and} \\
\langle N^2 \rangle &=& {1 \over 4} (\tilde Q, \tilde Q) \>T.
\eea
So the problem is to choose the function $\tilde Q$ which gives the
largest possible value of
\be \label{e:whattomax}
\left( {S \over N} \right)^2 = {\langle S \rangle^2 \over \langle N^2
\rangle} = \left( {3 H_0^2 \over 10 \pi^2 } \right)^2 T { (\tilde
Q,{\gamma(|f|) \Omega_{\rm gw}(|f|) \over |f|^3 P_1(|f|) P_2(|f|)})^2
\over (\tilde Q,\tilde Q)}.
\ee
But this is trivial!  Suppose you were given a fixed three-vector $\vec
A$, and asked to find the three-vector $\vec Q$ which maximized the
ratio ${(\vec A \cdot \vec Q)^2 \over \vec Q \cdot \vec Q}$.  Since
this is just the squared cosine of the angle between the two vectors,
the ratio is obviously maximized by choosing $\vec A$ and $\vec Q$ to
point in the same direction!  The problem of maximizing
(\ref{e:whattomax}) is identical; the solution is to choose
\be \label{e:best}
\tilde Q(f) = {\gamma(|f|) \Omega_{\rm gw}(|f|) \over |f|^3 P_1(|f|)
P_2(|f|)}.
\ee
Of course the overall constant appearing in the normalization of $Q$ has
no significance, since it does not affect the signal-to-noise ratio
(\ref{e:whattomax}).   We have now found the form of the optimal filter
for a stochastic background search.

\begin{figure}
\centerline{
\epsfig{file=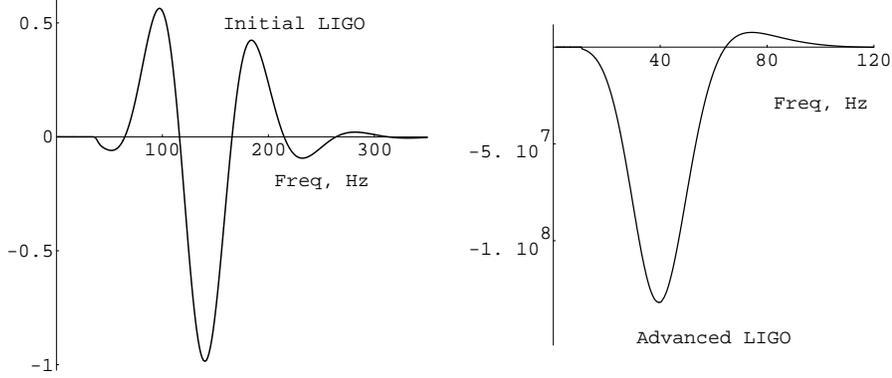,width=12cm,bbllx=75pt,bblly=298pt,bburx=533pt,bbury=490pt}}
\caption{\label{f:filters}
Optimal filter functions $\tilde Q(f)$ for the initial and advanced
LIGO detectors, for $\Omega_{\rm gw} =\rm constant$.  In this case
$\tilde Q(f)$ is proportional to $ {\gamma(f) \over f^3 P^2(f)}$.  The
initial LIGO filter is normalized to have maximum magnitude of unity;
the advanced LIGO filter is shown with an identical scale.  It is much
larger in magnitude because the power spectral noise density $P(f)$ of
the advanced detectors is much smaller than in the initial detectors.}
\end{figure}

These optimal filters are shown in Fig.~\ref{f:filters}
for both the initial and advanced LIGO detectors, for a flat
spectrum $\Omega_{\rm gw} = \rm constant$.
Effectively, the filter includes into the signal only those frequencies
at which the detector is most sensitive, weighting each frequency by a
positive or negative factor which reflects the relative phasing (at
that frequency) of the signals at the two detectors.

Having now found the optimal choice of filter function, it is
straightforward to calculate the signal-to-noise ratios for a given pair
of detectors.  Substituting the optimal choice of $Q$ given by
(\ref{e:best}) into
(\ref{e:whattomax}) gives the optimal signal-to-noise ratio:
\be \label{e:sovern}
\left( {S \over N} \right)^2 =
{9 H_0^4 \over 50 \pi^4} T \int_0^\infty df \> 
{\gamma^2 (f) \Omega_{\rm gw}^2(f) \over f^6 P_1(f) P_2(f)}.
\ee
Thus the signal-to-noise ratio of a correlation experiment to detect a 
stochastic
gravitational wave background is determined by the following quantities.
\beau
H_0 & = & {\rm Hubble\ expansion\ rate,\ sec}^{-1}. \\
T & = & {\rm Integration\ time,\ sec}. \\
\gamma(f) & = & {\rm Overlap\ reduction\ function,\ dimensionless}.\\
\Omega_{\rm gw} & = & {\rm Fractional\ energy\ density\ in\ gravitational\ waves}.\\
P_i & = & {\rm Noise\ power\ spectral\ density\ of\ detector\ {\it i}\ (one\ sided),\ sec}.
\eeau
One of the curious things about this formula for the signal-to-noise
ratio is that it depends upon the spectrum of gravitational waves
$\Omega_{\rm gw}$; a function which we do not know!  What this means in
practice is that rather than having a single ``optimal filter" one
needs to have a set of such filters.  The bandwidth of the experiment
is in a good sense defined by the factor $1 \over P_1(f) P_2(f)$
appearing in the integrand of (\ref{e:sovern}); the bandwidth of the
experiment is the range of frequencies over which this factor is large
(and hence the noise is small).  The signal-to-noise ratio gets its
significant contributions from within this bandwidth.  Within this
range of frequencies, which will not be very wide for the first
generation of detectors, it is reasonable to assume that the form of
the $\Omega_{\rm gw}$ is a power-law, proportional to $f^\alpha$.
Thus, one can construct a reasonable set of optimal filters by choosing
a number of different values of $\alpha$ such as
$\alpha=-4,-7/2,\cdots,7/2,4$ and analyzing the signal separately for
each filter.  The filter for $\alpha=0$ is shown in
Fig.~\ref{f:filters}.

In the next section, we will look at the values of this signal-to-noise
ratio for the forthcoming detectors, and estimate the detectable levels
of $\Omega_{\rm gw}$.

\subsection{
Signal to noise ratios, and the minimum
detectable $\Omega_{\rm gw}$
}
\label{s:ston} 
One can see from (\ref{e:sig2}) that in a measurement over the time
interval $T$ the signal $S$ is a sum (over $f$ and $f'$) of many
independent random variables (the Fourier amplitudes of the signals).
This is because the amplitudes $\tilde s_1(f)$ and $\tilde s_1(f')$ are
only correlated when $|f-f'| < 1/T \approx 10^{-7} \> \rm Hz$, and the
bandwidth over which the integral in (\ref{e:sig2}) gets its major
contribution is $\approx 100 \rm \>Hz$ wide.  Thus $S$ is the sum of
$\approx 10^{9}$ random variables, and the value of $S$ in a set of
measurements over independent time intervals is a Gaussian random
variable.  If the signal $S$ is measured many times, each time over a
distinct period of length $T$, the values of $S$ will have a Gaussian
normal distribution.  The mean value of this distribution is $\langle S
\rangle$; the width or standard deviation of this distribution away
from its mean value is $\langle N^2 \rangle^{1 \over 2}$.

In order to reliably detect a stochastic background, we need to be able
to state, with some certainty, that the value of $\Omega_{\rm gw}$ is
greater than zero.  Equivalently, one needs to be able to state with
some desired reliability that the observed positive mean value of
$S$ could not have resulted from detector noise but instead must have
resulted from a stochastic background.

One way of doing this is to compute the probability that a random variable
with a Gaussian normal distribution will lie within a given range.   For
example, the probability that a Gaussian random variable $S$ with
mean $\langle S \rangle$ and ${\rm variance}^2=\langle N^2 \rangle$
will lie within the range $[S_l,S_u]$ is
\be
{\rm Probability} = {1 \over \sqrt{2 \pi \langle N^2 \rangle}} 
\int_{S_l}^{S_u} dx \> \exp \left(- (x-<S>)^2 \over 2 \langle  N^2 \rangle \right) .
\ee
This integral can be expressed in closed form in terms of the error function
$\rm Erf$.  In this way, one can easily show that
\beau
{\rm with\ } 68\% {\rm \ probability\ }& S  \in & [ \ms-\sn2 , \ms+\sn2]\\
{\rm with\ } 90\% {\rm \ probability\ }& S  \in & [ \ms-1.65 \sn2 , \ms+1.65 \sn2]\\
{\rm with\ } 95\% {\rm \ probability\ }& S  \in & [ \ms-2 \sn2 , \ms+ 2\sn2]\\
{\rm with\ } 99.7\% {\rm \ probability\ }& S  \in & [\ms-3\sn2 , \ms+ 3\sn2].
\eeau
Thus, to detect a stochastic background with $ \ge 90\%$ confidence, we need
to ensure that the signal-to-noise ratio 
$\left( {S \over N} \right) \ge 1.65$, so that there is less than a ten percent
probability that $S$ is less than $0$.

As an example, let's analyze the sensitivity of the first and second
generation of LIGO detectors, to a flat spectrum $\Omega_{\rm gw}
\propto f^0$.  The design goals \cite{science92} for the noise power
spectral densities of the detectors are shown in Fig.~\ref{f:noise}.
The minimum flat spectral density detectable in $T=4 \rm \ months$ of
integration, with 90\% confidence, is given by
\bea
\Omega_{\rm  gw}^{\rm 90\%\ confidence}
&=& 1.65 \left[ {9 H_0^4 \over 50 \pi^4} T \int_0^\infty df \> 
{\gamma^2 (f) \over f^6 P_1(f) P_2(f)} \right]^{- {1 \over 2}}\\
& = & 2.8 \times 10^{-6} h_{100}^{-2} {\rm \ for\ initial\ LIGO}\\
& = & 2.8 \times 10^{-11} h_{100}^{-2} {\rm \ for\ advanced\ LIGO}.
\eea
We will see in the coming lectures that this is a very useful level of
sensitivity, particularly for the advanced detectors.

\begin{figure}
\centerline{\epsfig{file=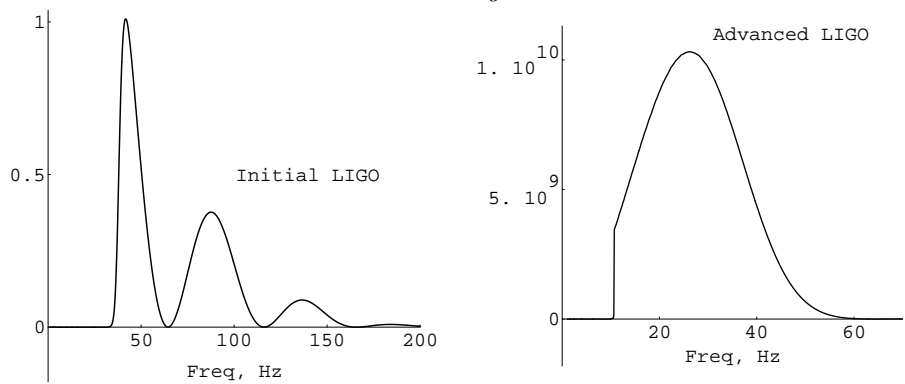, width=12cm, bbllx=80pt, bblly=298pt,
bburx=533pt, bbury=490pt}}
\caption{ \label{f:expected}
The expected ``signal" as a function of frequency for a flat spectrum
$\Omega_{\rm gw}=\rm constant$ in the initial and advanced LIGO
detectors.   The function displayed here is the integrand of
Eqn. \ref{eq30} ; it is proportional to ${\gamma^2(f) \over f^6
P^2(f)}$.  The initial LIGO signal is normalized to have maximum
magnitude of unity; the advanced LIGO signal is shown with an identical
scale.  It is much larger in magnitude because the power spectral noise
density of the advanced detectors is much smaller than in the initial
detectors.}
\end{figure}

One might wonder why the upgrade from the initial LIGO detectors to the
advanced detectors results in such a large increase in sensitivity.
The reason can be easily seen by comparing the optimal filter functions
$\tilde Q(f)$ for the initial and advanced detectors shown in
Fig.~\ref{f:filters}, and the expected contributions to $\langle S
\rangle$ given by (\ref{e:sig3}), as a function of frequency, shown in
Fig.~\ref{f:expected}.  It is easy to see that because the advanced
version of the LIGO detector will have much less low-frequency noise
than the initial detector, the expected signal in the LIGO detector
lies within the $0 < f < 64 \>\rm Hz$ region where the overlap
reduction function $\gamma(f)$ has the greatest magnitude.

\section{
What is known (experimental/observational
facts) about the stochastic background?}
\setcounter{equation}0
\label{s:what} 
So far in these lectures, we have discussed notation which may be used
to describe the spectrum of a stochastic gravitational wave background,
and some of the expected properties that such a background might have.
We also discussed, in detail, the optimal way to correlate the output
of two interferometers to search for such a background.

In this section, I'd like to show you what is actually known about the
stochastic background: the observational facts.  As you will see, there
are strict limits on the stochastic background in only a couple of
frequency ranges, but other than that, only one very general
constraint.  In particular, there are really no strong constraints on
$\Omega$ in the frequency range of interest for ground based
interferometric experiments or bar detectors ($1 \> {\rm Hz} < f < 10^3
\> {\rm Hz}$) or space-based experiments ($10^{-4} \> {\rm Hz} < f <
10^{-1} \> {\rm Hz}$).

\subsection{CMBR isotropy and the gravitational stochastic background}
\label{s:cobeconstraint}
The strongest constraint on $\Omega$ comes from the high degree of
isotropy observed in the Cosmic Microwave Background Radiation (CMBR).
In particular, the one-year 
(Smoot {\sl et al.} 1992; Bennett {\sl et al.} 1992; 
Wright 1992) and two-year data-sets
\cite{cobe2} from the Cosmic Background Explorer (COBE) satellite place
very strong restrictions on $\Omega$ at very low frequencies.  The
four-year data has also appeared, although it is not yet in print
\cite{cobe4}.

The temperature fluctuation $\delta T$ of the CMBR away from its mean
value of $T=2.73 \>\rm K$ varies from point to point on the sky, and may
be expanded in harmonic functions on the celestial sphere:
\be
{\delta T \over T}(\hat \Omega) = \sum_{lm} a_{l,m} Y_{lm}(\hat \Omega).
\ee
Here $\hat \Omega$ denotes a point on the two-sphere, and the $a_{l,m}$
are the multipole moments of the CMBR.   Since by definition the mean
value of the l.h.s. vanishes, the sum begins at $l=1$:
\be
\sum_{lm} \equiv \sum_{l=1}^\infty \sum_{m=-l}^l.
\ee
The COBE data permitted a determination of the rms quadrupole moment
of these fluctuations,
\be
Q_{\rm rms} \equiv T \sqrt{ \sum_{m=-2}^2 {|a_{2,m}|^2 \over 4 \pi} } =
14.3^{+5.2}_{-3.3} \> \mu \rm K.
\ee
These in turn place rather tight constraints on the spectrum of gravitational
radiation at very long wavelengths.

The connection between the temperature fluctuations of the CMBR and
gravitational waves arises through the {\it Sachs Wolfe effect}.  
Sachs and Wolfe (1967) showed how variations in the
density of the cosmological fluid and gravitational wave perturbations
result in CMBR temperature fluctuations, even if the surface of last
scattering was perfectly uniform in temperature.  In particular, they
showed that to lowest order the temperature perturbation at a particular
point on the sky is given by
\be 
{\delta T \over T}(\hat \Omega) = {1 \over 2} 
\int_{\rm null \> geodesic} d\lambda
{\partial \over \partial \eta} h_{rr}.
\ee
Here the integral is taken over a null geodesic path which leaves our
current spacetime point heading off in the spatial direction defined by
$\hat \Omega$ and going back to the surface of last scattering of the
CMBR, and $\lambda$ is a particular choice of affine parameter along
that null geodesic.  In conformal coordinates the metric perturbation
is given by $\delta g_{ij} = a^2(\eta) h_{ij}$, and $r$ is an outwards
radial spatial coordinate from our current spacetime point.  The effect
of a long wavelength gravitational wave is to shift the temperature
distribution of the CMBR on the celestial sphere away from perfect
isotropy.  The fact that the temperature perturbations are quite small
(a few parts in $10^6$) implies that the perturbations caused by the
gravitational waves can not be too large. (One specific example of this
effect can be found in 
Allen \& Koranda (1994); Koranda \& Allen (1995). In those
papers, we give a rigorous calculation of the effects of a stochastic
gravity-wave background on the CMBR temperature distribution, in an
inflationary model of the early universe.)

If we combine the COBE observational limits on the different
multipole moments $(2 \le l \le 30)$ we arrive at the
constraint on $\Omega_{\rm gw}$ given by
\be 
\Omega_{\rm gw} h_{100}^2 < 7 \times 10^{-11} \left({H_0
\over f} \right)^2 \quad \rm for \> H_0 < f < 30 H_0.
\ee
Note that this limit does not apply to {\it any} gravitational waves,
but only to those of cosmological origin, which were already present at
the time of last scattering of the CMBR.  This limit applies only over
a narrow band of frequencies at the very lowest limits of
Fig.~\ref{f:cmbr}; it is far below any frequencies that will be
accessible to investigation either by space-based or earth-based
experiments.

In Section~\ref{s:inflation} we will make a detailed examination
of inflationary cosmological models, and investigate these limits on
$\Omega_{\rm gw}$ in more detail.

\subsection{
Millisecond pulsar timing constraints on $\Omega_{\rm gw}$}
\label{s:pulsar}
For about a decade, Taylor and his research group \cite{Taylor}
have been monitoring the radio pulses arriving from a number of stable
millisecond pulsars.  Two of these pulsars are described in
Table~\ref{t:pulsars}.  These objects have turned out to be remarkably
stable clocks; the regularity of their pulses places tight constraints
on $\Omega_{\rm gw}$ at frequencies of order $1/{\rm observation \>
time} \approx 10^{-8} \rm Hz$.

\begin{table}
\caption{ Two of the millisecond pulsars studied by Taylor's group, and
then used to place limits on the stochastic gravitational wave
background. }
\hrule
\begin{center}
\begin{tabular}{cccc} 
Pulsar Name & Time Observed $t$ &  Period &  Timing Residuals $\Delta t$ \\
PSR B1855+09 & 7 Years & 5.4 msec & no residuals \\
PSR B1937+21 & 8 Years & 1.6 msec & few $\mu$sec (total) \\ 
\end{tabular}
\end{center}
\label{t:pulsars}
\end{table}

The regularity of these pulsar's pulses can be described in terms of
{\it timing residuals}.  Roughly speaking, after fitting the arrival
times of the pulses to a uniformly-spaced set (in the rest frame of the
pulsars) the timing residuals are the remaining errors in the emission
times of the pulses.  As can be seen from Table~\ref{t:pulsars} the
timing residuals are remarkably small; in dimensionless terms,
\be
{\Delta t \over t} \approx {\mu \rm sec \over 8 \> years} < 10^{-14}.
\ee
It is easy to see that these extremely precise measurements put very
tight restrictions on the amplitude of any gravitational waves with
periods comparable to 8 years.

To see the exact connection,
imagine that a gravitational wave is passing by us, coming from a
direction transverse to our line-of-sight to the pulsar, and assume its
wavelength is 8 light years.  Assume in addition that the pulsar is an
ideal clock, emitting pulses at perfectly regular intervals in its
rest frame.  Then, during the time when one of the maxima of the
gravitational wave is passing by the earth, the pulsar pulses will
arrive spaced slightly farther {\it apart} (you can think of this as
gravitational redshifting due to the gravitational wave).  Four years
later, a minimum of the gravitational wave will be passing by the
earth, and the pulsar pulses arrive at an earth-bound observer slightly
more closely spaced {\it together} (which can be thought of as
gravitational blueshifting).  In fact this system is just like a
single-arm gravitational wave detector, with an arm length $L$ of 8
light years.   The size of the characteristic gravity-wave strain must
be less than of order
\be
h_c \le \Delta L/L = \Delta t/t \le 10^{-14},
\ee
where $L=c t=8$ years and $\Delta t \approx \mu\rm sec $ is an upper
bound on the residual timing error.  Substituting $f =(8 \> \rm
years)^{-1} = 10^{-8}$ Hz in Eqn. (\ref{e:chirpamp}) gives the relation
$h_{\rm c} = 10^{-10} \sqrt{\Omega(f=10^{-8}{\rm Hz})} $ between the
characteristic strain and the fractional energy density in
gravitational waves.  Hence the extremely small size of the fractional
timing errors restricts $h_{\rm c}$ to be extremely small, which in
turn gives a bound on $\Omega$ of
\be
\Omega(f=10^{-8}{\rm Hz}) < 10^{-8}.
\ee
This bound provides important constraints on cosmic string networks, as
discussed in more detail in Section~\ref{s:cosmicstrings}.  However it
is not directly relevant for LIGO and other ground-based detectors,
because the bound is at a frequency 10 orders of magnitude lower than
their band of observation frequencies.

The sensitivity of these pulsar timing experiments is truly
remarkable!  For example, the authors point out that in their 7-year
data set for pulsar PSR B1855+09, the pulsar rotated precisely
40,879,349,533 times on its axis!  They also conclude that remaining
timing residuals are not due to gravitational waves: ``We do not
consider it likely that the observed trends in the timing residuals are
in fact caused by gravity waves."

I should note that a re-analysis of this pulsar timing data has
recently appeared \cite{thorsett}, which gives even more stringent
limits (at 90\% confidence) of $\Omega_{\rm gw}< 4.8 \times 10^{-9}$ at
$4.4 \times 10^{-9}$ Hz.  However this analysis has been criticized in
other work \cite{mchugh} which uses a different method (Bayesian rather
than frequentist Neyman-Pearson tests) to analyze the data, so it is
still not entirely certain that the more stringent limit is reliable.

\subsection{ Nucleosynthesis limits on $\Omega_{\rm gw}$}
\label{s:nucleo} 
The final bound on $\Omega$ is obtained from the standard model of
big-bang nucleosynthesis (often referred to as BBN \cite{kolbturner}).
This model provides remarkably accurate fits to the observed abundances
of the light elements in the universe, tightly constraining a number of
key cosmological parameters.  One of the parameters which is
constrained in this way is the expansion rate of the universe at the
time of nucleosynthesis.  This in turn places a constraint on the
energy-density of the universe at that time, which in turn constrains
the energy-density in a cosmological background of gravitational
radiation.

The BBN model's constraint can be expressed in terms of the number of
massless neutrinos, which can be present and in thermal equilibrium at
the time of nucleosynthesis.  One typically obtains a bound of the form
\cite{kolbturner}
\be
\int d(\ln f) \> \Omega_{\rm gw \> at \> nucleosynthesis} \le
{{7 \over 8} (N_\nu-3) \over 1+3 \times {7 \over 8} + 2 \times {7 \over 8}}
\left({\rho_{\rm rad} \over \rho_{\rm crit}}\right)_{\rm at \> nucleosynthesis}
\ee
where the quantities appearing on each side are the fractional energy
densities in gravitational radiation and in relativistic matter {\it at
the time of nucleosynthesis}. 
The integral includes only frequencies $f$ greater than the Hubble
expansion rate at the time of nucleosynthesis, in other words wavelengths
smaller than the Hubble radius at that time.
The BBN model constraints can be
expressed in terms of the number of massless neutrinos permitted;
this is $N_{\nu} \le 3.4$.    
To give a rough idea of the limits on the present-day spectrum of
$\Omega_{\rm gw}$, we can convert this limit at the time of nucleosynthesis
to a limit today.  Since the energy-density of radiation (relative to
the dominant energy density in the universe) redshifts $\propto (1+Z)^{-1}$
after the time of equal matter and radiation energy densities, this
inserts a factor of $(1+Z_{\rm eq})^{-1}$ into the limits.
One obtains a bound on $\Omega_{\rm gw}$
today of
\be
\int \Omega_{\rm gw } d(\ln f) \le 0.07 \times (1+Z_{\rm eq})^{-1}
\approx 10^{-5},
\ee
where $Z_{\rm eq} \approx 6000$ is the redshift at the time of equal
matter and radiation energy densities.  The range of integration now
includes only wavelengths shorter than about 10 pc.  This bound
restricts the spectrum of gravitational radiation $\Omega_{\rm gw}$
over a broad range of frequencies, but is not a very restrictive
constraint.

\section{Cosmological sources of a stochastic
background}
\setcounter{equation}0
\label{s:sources} 
In this section, we will examine in detail three different physical
processes which might have taken place in the early universe, which
give rise to different spectra of $\Omega_{\rm gw}$.  In addition to
their general interest, these three examples serve to motivate a number
of interesting observations and comments.

The first model we examine is the standard ``inflationary" scenario for
the early universe (Kolb \& Turner 1990; Linde 1990). 
 In this model, as the
universe cooled, it passed through a phase in which the energy-density
of the universe was dominated by vacuum energy, and the scale factor
increased exponentially-rapidly.  You will see that in its simplest
version, this scenario is tightly constrained by the CMBR observational
data 
(Allen \& Koranda 1994; Koranda \& Allen 1995). 
When this constraint is taken
into account, the stochastic gravity wave background predicted by the
simplest inflationary cosmological models are far too weak to be observable
by either the initial or advanced LIGO detectors.

The second model we examine is the cosmic string scenario
(Vilenkin 1985; Vilenkin \& Shellard 1994).  In this model, as the
universe cooled, long stringlike defects were formed at a phase
transition.  These ``cosmic strings" form a network which
self-intersects and chops off small loops of string.  These small loops
oscillate relativistically, and as they do so they emit a
characteristic spectrum of gravitational waves.  In contrast with the
previous example, for reasonable values of the parameters, the cosmic
string scenario predicts a stochastic background which is large enough
to be observable with the advanced LIGO detector.  In common with the
previous example, the string network is described by a ``scaling"
solution, with the consequence that the spectrum of $\Omega_{\rm gw}$
is flat over a wide range of frequencies.

The third and final example in this section is also motivated by a
phase transition in the early universe.  We consider the case in which
the phase transition is strongly first-order, and creates vacuum
bubbles, which are bubbles of the new (low energy density) phase
expanding in the old (high energy density) phase
(Kamionkowski {\sl et al.} 1994; Kosowsky {\sl et al.} 1992; 
Kosowsky \& Turner 1993).  Within a short time, these bubble
walls are moving relativistically, and much of the latent energy
released by the phase transition is transformed into kinetic energy of
the bubble walls. The highly-relativistic collisions of these bubble
walls are a copious source of gravitational radiation, strong enough to
be observable (in some cases) with the initial generation of LIGO
detectors.  Unlike the previous two examples, this phase
transition/gravitational radiation production process is a ``one-time"
event, characterized by a particular cosmological time and a particular
frequency today.  Hence, in contrast with the previous two examples,
you will see that the spectrum $\Omega_{\rm gw}(f)$ is strongly peaked
at a frequency characteristic of this time.

These three examples are intended primarily to be illustrative.  They
are by no means comprehensive.  The fact of the matter is that we don't
know what happened in the early universe -- the subject is extremely
speculative, and will remain so until we have data (such as the
spectrum of $\Omega_{\rm gw}$) that carries specific information about
very early times.   Nevertheless, these examples are quite useful,
because they illustrate some of the mechanisms that might produce a
cosmological stochastic background.   They certainly lead me to suspect
that such a background is present, and to hope that it might be
observable.

\subsection{Inflationary cosmological models
}
\label{s:inflation} 
Inflationary models of the early universe were studied extensively in
the early- and mid-1980's (for a review, see Linde (1990)). Generally
speaking, these are a class of cosmological models in which the
universe undergoes a phase of very rapid (power-law or exponential in
time) expansion at early times.  Inflationary models have a number of
nice properties.  They are very simple, and highly predictive.  They
provide ``natural" solutions to the horizon, flatness, and monopole
problems.  They are also in good agreement with the COBE observations
of the spectrum of temperature perturbations in the CMBR.  (Note,
however, that many other types of early-universe models are also in
good agreement with these observations!)

One of the remarkable facts about inflationary models is that they
contain a beautiful mechanism which creates perturbations in all
fields.   These provide, within the context of inflationary models, a
natural mechanism to create the density perturbations which evolved to
form galaxies and clusters of galaxies today.  More important for our
purposes is that this mechanism also gives rise to an extremely
distinctive spectrum of stochastic gravitational radiation.

In this lecture, I will show you how these perturbations in
inflationary cosmologies arise from the most basic quantum mechanical
effect: the uncertainty principle.  In effect, the spectrum of
gravitational radiation that might be observed today is nothing other
than adiabatically-amplified zero-point fluctuations.

\begin{figure}
\centerline{\epsfig{file=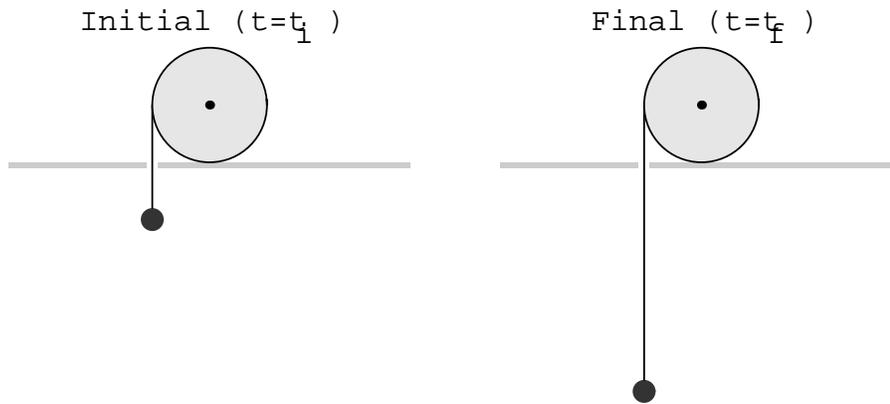, width=12cm, bbllx=90pt, bblly=310pt,
bburx=520pt, bbury=504pt}}
\caption{ \label{f:oscillator}
The pendulum consists of a mass hanging from a string.  Initially the
string is short, and the oscillation frequency is large.  Between
initial and final times $t_i$ and $t_f$, the string is reeled out,
increasing its length and decreasing the oscillation frequency of the
mass.  The quantum mechanical behavior of this system is more
interesting than its classical behavior, and provides a simple model
for the mechanism that generates large classical perturbations from
zero-point fluctuations in inflationary cosmological models.}
\end{figure}

Let me begin with a simple example.  Suppose that we have a harmonic
oscillator like the one shown in Fig.~\ref{f:oscillator}.  It is a
traditional pendulum, consisting of a mass on a string.  In this case
however the string is wound around a reel, and by unwinding the reel
one can increase the length of the string and decrease the oscillation
frequency $\omega = \sqrt{g/l}$ of the system.  Here $l$ is the length
of the string and $g$ is the acceleration of gravity at the surface of
the earth.  We will denote the initial frequency by
$\omega_i=\sqrt{g/l_i}$ and the final frequency by
$\omega_f=\sqrt{g/l_f}$. Let us suppose that the mass is initially at
rest in its equilibrium position and consider in turn its classical and
its quantum behavior.

The {\it classical} behavior of this system is not very exciting.
Since the mass is at rest in its minimum energy configuration (hanging
straight down) it simply remains there.  After reeling out the string,
and lowering the frequency of the oscillator, the mass remains at
rest, hanging straight down.

The {\it quantum mechanical} behavior of this system is far more
interesting.  It has a long history (neatly encapsulated as the
``adiabatic theorem") dating back to the invention of quantum mechanics
in the 1920's.  Because of the uncertainty principle, it is impossible
for the oscillator to be truly at rest; in the minimum energy state it
has initial energy $E_i = {1 \over 2} \hbar \omega_i$, and executes
zero-point motion about its classical rest position.  In this initial
state, the initial number of quanta $ N_i  =0$.  Quantum
mechanically, this system has two possible types of behavior, depending
upon how the string is reeled out.  To distinguish these two cases, let
$T=t_f-t_i$ denote the time over which the string was steadily reeled
out.

The case where $T$ is quite long compared to both the initial and final
oscillation periods, $T>>2 \pi/\omega_f$, is the slow or adiabatic
case.  In this case, the final energy of the oscillator is $E_f={1
\over 2} \hbar \omega_f$.  The difference between this energy and the
initial energy represents work done to the reel that lowered the
string.   In addition to the ``classical" work $mgh$ done to the
friction mechanism of the reel, a bit of additional ``quantum" work
$E_i-E_f$ is also done.  In this case, no quanta are created, $
N_f =0$ and the mass is still in its ground state after
being lowered down.

The truly fascinating case is the one in which $T$ is quite short
compared to the oscillation periods, $T<<2 \pi/\omega_i$.  (Strictly
speaking, we need $\dot l / l > \omega $ at every moment.) In this
case, the behavior of this quantum mechanical oscillator is quite
different than the behavior of the corresponding classical oscillator.
In this case, as I will shortly show you, the final energy is {\it
half} of the initial energy:  $E_f = {1 \over 2} E_i = {1 \over 4}
\hbar \omega_i$.  In this case, we create quanta: $ N_f = {1 \over 4}
{\omega_i \over \omega_f}$.  It is in this way that the inflationary
universe models create large macroscopic fluctuations today, starting
only with the minimum fluctuations expected from the uncertainty
principle.  In order to solve the horizon and flatness problems,
inflationary models require the universe to expand by a factor greater
than $\approx 10^{27}$.  This redshifts frequencies by the same amount,
so that the ratio ${\omega_i \over \omega_f} > 10^{27}$ and one ends up
with large occupation numbers at the present.

\begin{figure}
\centerline{\epsfig{file=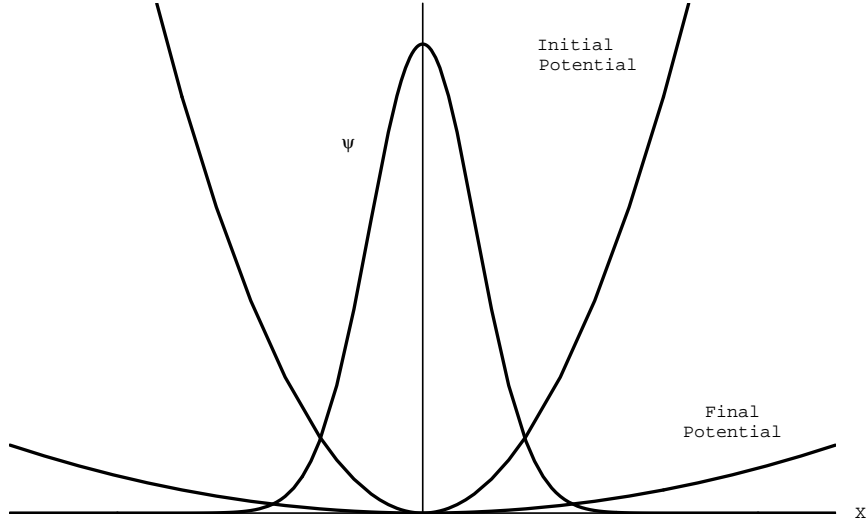, width=12cm, bbllx=60pt, bblly=250pt,
bburx=555pt, bbury=540pt}}
\caption{ \label{f:adiabatic}
A particle moving in the initial harmonic oscillator potential is in
the ground state, and has a wavefunction shown by $\psi(x)$.  The
frequency of the oscillator is suddenly decreased, by external means,
and the potential becomes that shown as the final potential.  The
wavefunction no longer corresponds to the ground state, but to a
highly-excited many-particle state.}
\end{figure}

Before going into the details of inflation, let me show you a simple
calculation to justify and explain the claims made in the previous paragraph.
Shown in Fig.~\ref{f:adiabatic} are the initial and final potentials
of the harmonic oscillator pictured previously in Fig.~\ref{f:oscillator}.
Let $\alpha_i=m \omega_i/\hbar$ where $m$ is the mass of the ball.
The initial wavefunction corresponding to the ground state is
\be
\psi(x)=\left({\alpha_i \over \pi } \right)^{1/4}
e^{- { \alpha_i \over 2} x^2}.
\ee
Imagine now that at time $t=0$ this potential is suddenly changed to a
final potential which is another harmonic oscillator potential but with
much lower frequency $\omega_f$, as shown in Fig.~\ref{f:adiabatic}.
Let $\alpha_f=m \omega_f/\hbar$.  Then we can express the
wavefunction at time $t=0$ in terms of the energy eigenstates of the
new harmonic oscillator potential:
\be \label{e:decompose}
\psi(x) = \sum_{n=0}^\infty c_n \psi_n(x)
\ee
where $\psi_n(x)$ are the energy eigenfunctions of the final (lower frequency)
potential:
\be
\psi_n(x) = N_n H_n(\sqrt{\alpha_f} x) e^{- { \alpha_f \over 2} x^2}.
\ee
Here $N_n$ is a normalization constant, and the $H_n$ are Hermite polynomials.
In the usual way, we can work out the values of the expansion coefficients
$c_n$: we multiply both sides of (\ref{e:decompose}) by $\psi^*_m(x)$ and
integrate over all $x$, using the orthogonality of the eigenfunctions to
obtain a formula for $c_m$:
\bea
c_n & = & 0 {\rm \ \ for\ n\ odd,\ and}\\
c_{2n} & = & (\alpha_i \alpha_f)^{1/4} \sqrt{2 (2n)! \over \alpha_i + \alpha_f}
{1 \over n!} \left( {\omega_f - \omega_i \over 2(\omega_f + \omega_i)} \right)^n 
\eea
It is now straightforward to calculate the expectation value of the energy:
\be
E_f = \sum_{n=0}^\infty (2n+ 1/2) \hbar \omega_f | c_{2n}|^2 = \hbar
\omega_f \left({1 \over 2} + {(\omega_f-\omega_i)^2 \over 4 \omega_i
\omega_f } \right).
\ee
It is easy to see that in the limit when the final frequency is much lower
than the initial frequency, $\omega_f<<\omega_i$, one obtains
$E_f \approx {1 \over 4} \hbar \omega_i = \hbar \omega_f
\left(  {1 \over 4} {\omega_i \over \omega_f} \right).$
This allows us to identify the number of created quanta as
$N_f = \left(  {1 \over 4} {\omega_i \over \omega_f} \right).$

This simple mechanical model shows exactly how inflationary models
generate perturbations in a ``natural" way.  Every mode of every
quantum field can be thought of as a simple harmonic oscillator, and
when the universe expands exponentially rapidly in size, the
frequencies of these modes drop extremely rapidly.  As you will see
shortly, in inflationary models, the universe expands rapidly in size,
and because this expansion is very quick, it is {\it not adiabatic}!
In other words, it corresponds to our second case above, where quanta
are created.  Even though this process is strictly quantum mechanical,
and the energy created is proportional to $\hbar$, the occupation
numbers today are large, and these perturbations appear today to be
large macroscopic fluctuations.  In the case of the gravitational field
itself, this process creates a characteristic spectrum of stochastic
gravitational waves.  Let me now sketch the calculation for a simple
inflationary model.  The details of this calculation can be found in
Allen (1988).

\begin{figure}
\centerline{\epsfig{file=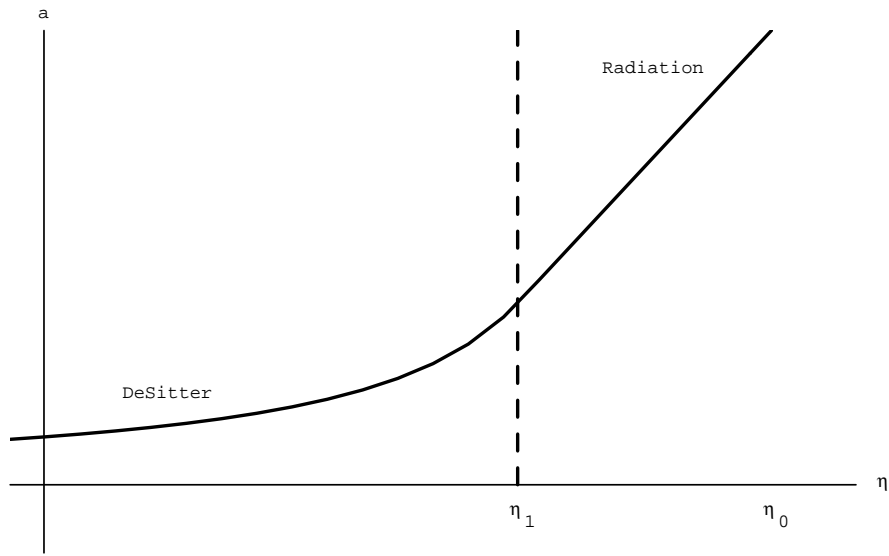, width=12cm, bbllx=72pt, bblly=246pt,
bburx=555pt, bbury=550pt}}
\caption{ \label{f:inflation}
The cosmological scale factor $a(\eta)$ as a function of conformal time
${\eta}$, for a simple inflationary model.  The universe is deSitter
space until time ${\eta}={\eta}_1$ when it makes an instantaneous phase
transition to a radiation-dominated epoch.  This would be a good
``crude" model of our own universe, if slightly to the past of the
present time ${\eta}_0$ the universe became matter-dominated and
$a$ increased somewhat more rapidly between that time and the
present.}
\end{figure}

We assume that the universe is described by a simple
``two-stage" cosmology as shown in Fig.~\ref{f:inflation}.
The metric of spacetime is given by
\be
ds^2 = a^2(\eta) ( -d\eta^2 + d \vec x^2 + h_{ab}(\eta,\vec x) dx^a dx^b),
\ee
where the scale factor $a(\eta)$, the Hubble expansion rate $H(\eta)$ and
the equation of state are given in Table~\ref{t:cosmology}.  (Note: $\eta$
is a conformal time coordinate with units of length.)
\begin{table}
\caption{ 
The cosmological scale factor $a(\eta)$ for a two-stage inflationary
cosmology.}
\begin{center}
\begin{tabular}{ccccc}
Time & Phase & Equation of State & Scale Factor $a({\eta})$ & Hubble $H({\eta})$ \\
${\eta}<{\eta}_1$ & DeSitter  & $-P=\rho=const.$ & ${\eta}_1^2 {\eta}_0^{-1} (2 {\eta}_1 - {\eta})^{-1}$ & $H_{\rm ds} = c {\eta}_0/{\eta}_1^2$ \\ 
${\eta}>{\eta}_1$ & Radiation & $ P=\rho/3$            &
${\eta}/{\eta}_0 $                          & $ c {\eta}_0/{\eta}^2$
\\
\end{tabular}
\end{center}
\label{t:cosmology}
\end{table}
Note that although the expansion does not appear to be very rapid
during the deSitter phase (${\eta}<{\eta}_1$) this is because the scale
factor $a({\eta})$ has been expressed in terms of a conformal time
coordinate.  When expressed in terms of physical or comoving time
defined by $ c dt= a({\eta}) d{\eta}$ the scale factor is $\propto
e^{H_{\rm ds} t}$ during the deSitter phase, and $\propto t^{1/2}$
during the radiation-dominated phase.  In order to solve the horizon and
flatness problems, enough expansion is needed to ensure that
$a(\eta_0)/a(\eta_1) > 10^{27}$ for GUT-scale inflation.  So typically
one has $\eta_0 > 10^{27} \eta_1$.

The gravitational waves in this simple two-stage cosmology are
described by metric perturbations of the form \cite{allen}
\be
h_{ab} = e_{ab}(\hat k) \phi({\eta}) \exp(i \vec k \cdot \vec x),
\ee
where $\vec k$ is a constant (in time $\eta$) wavevector.  The
polarization tensors $ e_{ab}$ are given in equations
(\ref{e:polar},\ref{e:polar2}) of Appendix~\ref{s:appendix1}.  The amplitude
$\phi$ satisfies a wave equation (the massless Klein-Gordon equation
\cite{allen})
\be
\left( {d^2 \over d{\eta}^2} + {2 \over a} {da \over d{\eta}} {d \over
d{\eta}} + |\vec k|^2 \right) \phi=0.
\ee
The solutions to this wave equation may be easily found in terms of elementary functions
(which are simple cases of half-integer Bessel or Hankel functions) in both the
deSitter and radiation-dominated epochs:
\bea \label{e:groundstate}
{\rm For\ } {\eta}< {\eta}_1 \quad \phi(\eta) & = & {a({\eta}_1) \over
a({\eta})} \left[ 1+i H_{\rm ds} \omega^{-1} \right]
e^{-i k({\eta}-{\eta}_1)} \\
{\rm For\ } {\eta}> {\eta}_1 \quad \phi(\eta) & = & {a({\eta}_1) \over
a({\eta})} \left[ \alpha e^{-i k({\eta}-{\eta}_1)}
+ \beta  e^{i k({\eta}-{\eta}_1)} \right],
\eea
where $\omega=c k/a$ is the angular frequency of the wave, $\alpha$ and
$\beta$ are time-independent constants.  Note that because $k=|\vec k|$
is constant, the angular frequency $\omega$ is a function of time.
Since this differential equation is second order, one may obtain the
values of the constants $\alpha$ and $\beta$ by demanding that both
$\phi$ and $d \phi/d{\eta}$ are continuous at the boundary
${\eta}={\eta}_1$ between the deSitter and radiation epochs of
expansion.  Imposing this restriction on the solution leads to
\bea
\alpha &=& 1 + i { \sqrt{H_0 H_{\rm ds}} \over \omega } - {H_0 H_{\rm
ds} \over 2 \omega^2}\\
\beta &=&  {H_0 H_{\rm ds} \over 2 \omega^2}.
\eea
In these expressions, we use $\omega = c k/a(\eta_0)$ to denote the
angular frequency that would be observed today, and $H_0 = c /{\eta}_0$
for the Hubble expansion rate that would be observed today.  In effect,
this problem is identical to tunneling problems done in elementary
quantum mechanics, and we are simply calculating
transmission/reflection coefficients.   These types of calculations are
also frequently referred to as Bogoliubov coefficient methods in the
literature \cite{birrelldavies}.

In an inflationary model, the long period of inflation damps out any
classical or macroscopic perturbations, leaving behind only the minimum
allowed level of fluctuation, required by the uncertainty principle.
We have chosen our wave-function (\ref{e:groundstate}) to correspond
precisely to this deSitter vacuum state.  Provided that the period of
inflation was long enough, the observable properties of the universe
today should be indistinguishable from those of a universe which
started in this vacuum state.  We could work out the correct overall
normalization of the wavefunction, but it is not necessary.

Today, in the radiation-dominated phase of the universe, the eigenmodes
describing particles are the coefficients of $\alpha$, and those
describing antiparticles are the coefficients of $\beta$.  One may
easily show that the number of created particles of angular frequency
$\omega$ today is given by
(Allen 1988; Birrell \& Davies 1982)
\be
N_\omega \equiv {\rm Number\ of\ created\ particles\ of\ freq\ } \omega
= |\beta_\omega|^2 =
{H_0^2 H_{\rm ds}^2 \over 4 \omega^4}.
\ee
Along the lines of the calculation given in Section~\ref{s:ssdig} and
in particular in analogy with (\ref{e:diff}), we may write an
expression for the stochastic gravitational-wave energy-density
contained in the (angular) frequency interval $(\omega,\omega+d\omega)$
as
\be
d\rho_{\rm gw} = 2 \hbar \omega \left( {\omega^2 d\omega \over 2 \pi^2
c^3} \right) N_\omega=
{\hbar H_0^2 H_{\rm ds}^2 \over 4 \pi^2 c^3} {d\omega \over \omega}=
{\hbar H_0^2 H_{\rm ds}^2 \over 4 \pi^2 c^3} {df \over f}.
\ee
This may be re-written in terms of the present-day energy-density of the universe,
and the (constant) energy-density during the deSitter phase, in the following way.
The Hubble expansion rates are given in terms of these energy-densities by
$H_0^2 = {8 \pi G \rho_c \over 3 c^2}$ and $H_{\rm ds}^2 = {8 \pi G \rho_{\rm ds} \over 3 c^2}$.
The spectrum of stochastic gravitational waves is thus
\be
\Omega_{\rm gw} = {f \over \rho_c} {d\rho_{\rm gw} \over df} = {16 \over 9} {
\hbar G^2 \over c^7} \rho_{\rm ds} = {16 \over 9} {\rho_{\rm ds} \over \rho_{\rm Planck}},
\ee
where we have introduced the Planck energy density 
\be
\rho_{\rm Planck} = {c^7 \over \hbar G^2}
= 10^{115} \rm \ ergs/cm^3.
\ee
You will notice that $\Omega_{\rm gw}$ is proportional to $\hbar$,
because this background is nothing other than
(parametrically-amplified) zero-point energy!

Several comments are now appropriate.  First, our calculation of $\Omega_{\rm gw}$ is
for a simplified cosmological model that does not include a matter-dominated phase after
the radiation-dominated epoch of expansion.  If this matter-dominated phase is included,
the result obtained is \cite{allen}
\be
\Omega_{\rm gw} (f)=  {16 \over 9} {\rho_{\rm ds} \over \rho_{\rm
Planck}} (1+Z_{\rm eq})^{-1},
\ee
for those waves which, at the time the universe became
matter-dominated, had a wavelength shorter than the Hubble length at
that time.  Today this corresponds to frequencies $f>(1+Z_{\rm
eq})^{1/2} H_0$, where $Z_{\rm eq}$ is the redshift of the universe
when the matter and radiation energy densities were equal.  At lower
frequencies, the spectrum of $\Omega_{\rm gw}(f) \propto f^{-2}$.
Second, note that our result (which is frequency-dependent) does not
truly apply ``from DC to light".  For waves which are long compared to
the Hubble length ($10^{28}$ cm today) corresponding to frequencies
less than $H_0$ today, the notion of energy density becomes
non-sensical, because the wavelength becomes longer than the curvature
length scale of the background space-time.  In similar fashion, at high
frequencies there is a maximum frequency above which $\Omega_{\rm gw}$
drops rapidly to zero.  In our calculation, we assume that the phase
transition from the deSitter to the radiation-dominated epoch is
instantaneous.  However this process occurs over some time scale
$\Delta \tau$, and above a frequency
\be
f_{\rm max} = {a(t_1) \over a(t_0)} { 1 \over \Delta \tau}
\ee
which is just the redshifted characteristic rate of the transition, $\Omega_{\rm gw}$ 
drops rapidly to zero.
These low- and high-frequency cutoffs to the spectrum guarantee that the total energy-density
in gravitational waves is finite rather than infinite.

\begin{figure}
\centerline{\epsfig{file=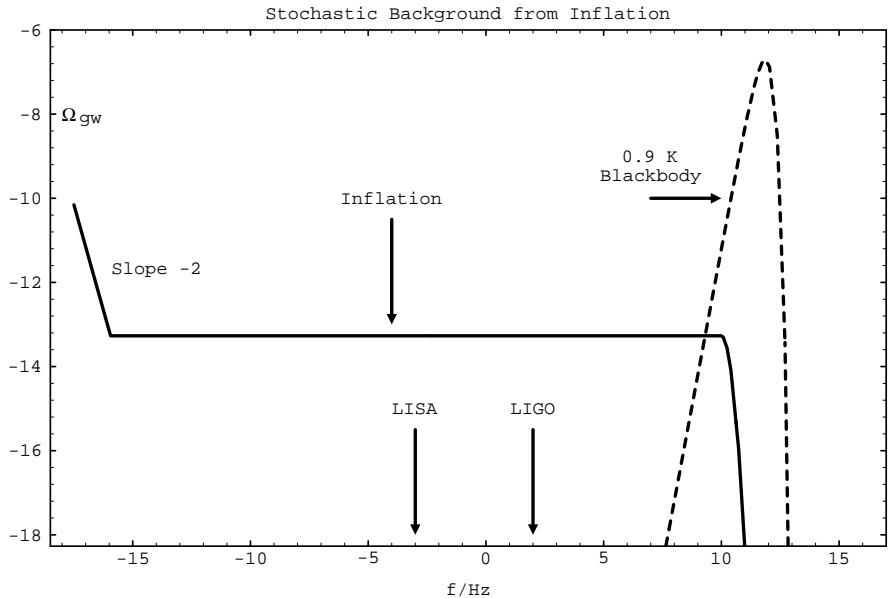, width=12cm, bbllx=72pt, bblly=240pt,
bburx=555pt, bbury=560pt}}
\caption{ \label{f:infspect}
The spectrum of stochastic gravitational waves in inflationary models
is flat over a wide range of frequencies, and is shown as the solid
curve.  The horizontal axis is $\log_{10}$ of frequency, in Hz.  The
vertical axis is $\log_{10} \Omega_{\rm gw}$.  The inflationary
spectrum rises rapidly at low frequencies (wave which re-entered the
Hubble sphere after the universe became matter-dominated) and falls off
above the (appropriately redshifted) frequency scale $f_{\rm max}$
associated with the fastest characteristic time of the phase transition
at the end of inflation.  The amplitude of the flat region depends only
on the energy-density during the inflationary stage; we have chosen
the largest amplitude consistent with the COBE constraint discussed
earlier: $\Omega_{\rm gw}<7 \times 10^{-11}$ at $10^{-18}$ Hz.  This
means that at LIGO and LISA frequencies, $\Omega_{\rm gw}< 8 \times
10^{-14}$.  Shown for comparison is the spectrum of a $0.9$ K
blackbody, as discussed earlier.}
\end{figure}

A typical spectrum of gravitational waves resulting from inflation is
shown in Fig.~\ref{f:infspect}.  For GUT energy-scale inflation, the
ratio
\be
{\rm GUT-scale\ Inflation:} \quad {\rho_{\rm ds} \over \rho_{\rm
Planck}} \approx (10^{16} \> {\rm GeV} / 10^{19} \> {\rm GeV})^4 \approx
10^{-12}.
\ee
(Also shown for comparison is the naive
0.9 K blackbody, discussed in Section~\ref{s:ssdig}.) We have also chosen
$f_{\rm max} \approx 10^{10} \> \rm Hz$, a reasonable value for
GUT-scale inflation.  You can see right away that in the inflationary
case, the constraint coming from the COBE observations discussed in
Section~\ref{s:cobeconstraint} puts very severe restrictions on the
spectrum (Allen \& Koranda 1994; Koranda \& Allen 1995).  
In Fig.~\ref{f:infspect} we
have chosen the amplitude of the spectrum (determined by the ratio $
{\rho_{\rm ds} \over \rho_{\rm Planck}}$) to be {\it as large as
possible, consistent with this COBE constraint}.  Nevertheless, because
the spectrum falls off $\propto f^{-2}$ at low frequencies, this means
that today, at LIGO and LISA frequencies (indicated by the arrows),
$\Omega_{\rm gw}< 8 \times 10^{-14}$.  This is certainly too small to
be observable with either initial or advanced LIGO, and is at the
limits of what might be potentially observable with LISA.

Recent work \cite{turner96} has examined in more detail the spectrum of
stochastic gravitational radiation produced in slow-roll inflationary
models.  In these models, the spectrum is not exactly flat at high
frequencies, but has a (negative) logarithmic slope $n_T$.  There is a
connection between this slope and the amplitude of the spectrum at the
lowest frequencies; if the amplitude is large then $n_T$ is quite
negative, and if $n_T$ is near zero then the amplitude is small.
Consequently Turner found that in slow-roll inflation the upper bounds
on $\Omega_{\rm gw}$ at LIGO and LISA frequencies are about an order of
magnitude {\it lower} than the ones given here.

There are other types of ``inflationary" cosmology based on superstring
models of the fundamental interactions, which have recently received a
great deal of attention \cite{brustein}.  These models typically
contain a ``deflationary" epoch followed by a stage of rapid expansion,
and produce a spectrum in which $\Omega_{\rm gw}(f) \propto f^3$ below
some maximum cutoff frequency.  This spectrum evades the CMBR and
millisecond pulsar timing constraints, yet might still produce enough
stochastic background to be observable by LIGO \cite{allenbrustein}.

Before moving on to our next example (cosmic strings) of how
cosmological processes can give rise to a stochastic background, a few
final comments are in order.  The first is that the long flat region of
the spectrum $\Omega_{\rm gw}$ reflects the scaling behavior of the
cosmological scale factor $a(t) \propto \exp(H_{\rm ds}
t)$ during the deSitter phase of expansion.  We will see
identical self-similar behavior in our next example.  Notice that the
highest frequency parts of the spectrum (such as the falloff at $10^{10}$
Hz in our example) reflect the behavior of the universe at early times
(in this example, the nature of the phase transition that ended
inflation at $\approx 10^{-35}$ seconds after the big bang).  The low
frequency parts of the spectrum reflect the fairly recent history of the
universe.  For example, for inflationary cosmology, the ``kink" in the
spectrum at $10^{-16}$ Hz reflects the change in the expansion law
from radiation- to matter-domination at $Z_{\rm eq} \approx 6000$.
Grishchuk has shown that in general, this type of mechanism (parametric
amplification of zero-point fluctuations) produces a spectrum of the
sort we have illustrated.  However if the universe does not expand
exponentially (constant $H(t)$), but ``speeds up" and ``slows down" (i.e.
$H(t)$ increases and decreases) then this is reflected in the spectrum
of $\Omega_{\rm gw}$ which is no longer flat, but instead has
characteristic increases and decreases, reflecting the changing values
of $H(t)$ during the evolution of the universe.

\subsection{Cosmic string cosmology}
\label{s:cosmicstrings} 
In this lecture, I am going to show you a calculation of the stochastic
gravitational wave background produced by a network of cosmic strings
(Vilenkin 1985; Vilenkin \& Shellard 1994). 
 These are objects which might have formed
during a phase transition, as the universe cooled.  We do not know if
our universe contains cosmic strings or not (this depends upon the
precise nature of the phase transition) so these calculations must be
viewed as illustrative rather than realistic.

Cosmic strings are one-dimensional (in space) string-like objects.
They are topologi\-cally-stable scalar/gauge field configurations, whose
dominant decay mechanism is the emission of gravitational
waves.  These string-like objects are analogous to the vortex lines
which form in superfluid helium, or to topological defects which can
form during phase transitions of liquid crystals.  Here, our interest
is in cosmic strings which might have formed during a phase transition
as the universe cooled, and a fundamental $U(1)$ local gauge symmetry
was broken.  These are the simplest strings; they are characterized by
a single dimensional scale: their mass-per-unit-length $\mu$.  A phase
transition which formed strings at the Grand Unified Theory (GUT)
energy scale of about $10^{16}$ GeV would result in strings with $\mu
\approx 10^{22} \> {\rm gms/cm}$.

These kinds of cosmic string have several remarkable properties.
First, they are formed without any ``ends" so the strings are always in
the form of loops.   Sometimes the loops are small (compared to the
Hubble length) in which case we call them ``loops".  Sometimes the
loops are large (compared to the Hubble length) in which case we call
them ``infinite strings".  In neither case are there any unjoined
``ends".  Note that in assigning a ``length" to the loop, we are using
an invariant measure of the length; the length is defined as the total
energy of the loop divided by its mass-per-unit-length $\mu$.  A second
remarkably property of these loops is that they have a tension equal to
their mass-per-unit-length $\mu$.  This is in contrast to the kind of
strings we use to wrap packages, which have small tension compared to
their mass-per-unit-length $\mu$.  Because of this large tension, the
cosmic strings oscillate relativistically under their own tension.
Thus, a circular loop of cosmic string, initially placed at rest, will
collapse under its own tension in a time comparable to the time it
takes a light ray to move a distance equal to the radius of the loop.
In general, a non-circular loop will oscillate quasi-periodically, with
a period equal to (half) the time it takes light to travel a distance
equal to the length of the string making up the loop.  A third
remarkable property of these cosmic string loops is that they are
stable against all types of decay apart  from the emission of
gravitational radiation.   So typically a loop of cosmic string, under
the effects of its own tension, will oscillate almost periodically,
gradually emitting gravitational radiation, and shrinking
in size.

\begin{figure}
\begin{center}
\epsfig{file=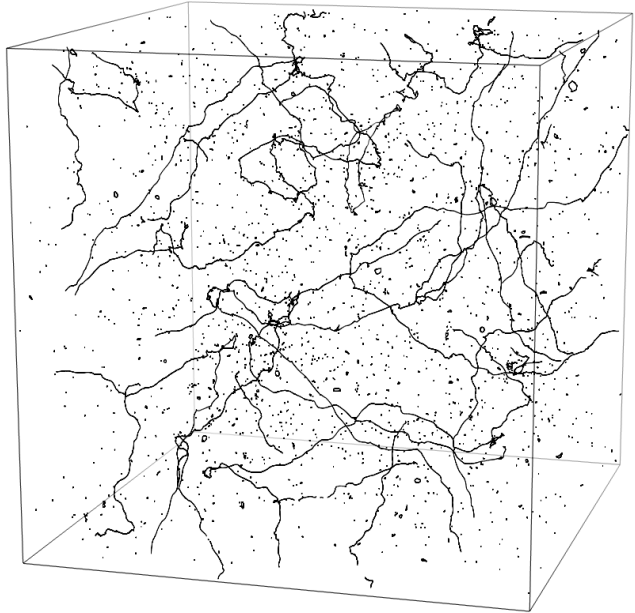, width=9cm, bbllx=82pt, bblly=214pt,
bburx=512pt, bbury=635pt}
\end{center}
\caption{ \label{f:strings}
A snapshot of a universe containing a cosmic string network.  The
spatial volume shown here is a cube with side length $2c /H$ where $H$
is the Hubble expansion rate.  The string network has a remarkable
scaling property: at any time the appearance of the network is
(statistically) like the one shown here, provided that all length
scales are expressed in terms of the Hubble length at that time.   The
network consists of many tiny loops, which here appear as points, and
a number of so-called ``infinite" strings.  Although these infinite
strings appear to have ends, this is just an artifact of having ``cut
out" a Hubble volume cube for this snapshot.}
\end{figure}

One might worry that the configuration of these cosmic strings is very
complicated, depending precisely upon how the mass-energy of the
universe was distributed when they were formed, and that for this
reason, it would be terribly difficult to make any precise predictions
about the effects they would have on the universe.   Fortunately, about
fifteen years of extensive study of networks of cosmic string has
revealed that they have a remarkable ``scaling" property
(Vilenkin 1985; Vilenkin \& Shellard 1994).   This can be summarized by
Fig.~\ref{f:strings}, which shows a snapshot (at one instant of cosmic
time) of a universe containing cosmic strings.
The crucial point is this one: at any instant in time, the appearance
of the string network is (statistically) like that shown in the figure
Fig.~\ref{f:strings}.  As the universe expands, the string network
evolves; however at any time, each Hubble sized volume contains, on the
average, a certain number (say A=52) of infinite strings passing through
it, and a very large number of small loops.  The small loops oscillate,
emitting gravitational radiation, and as the universe evolves, more
small loops are chopped off the infinite strings, replacing those loops
which are disappearing after converting their energy into gravitational
waves.   All of the properties of the network ``scale" with the Hubble
length.  So even though it is {\it impossible} to predict the appearance of a
string network in our own universe, in a statistical sense, we {\it can}
describe the average properties of the network at any time.

\begin{figure}
\begin{center}
\epsfig{file=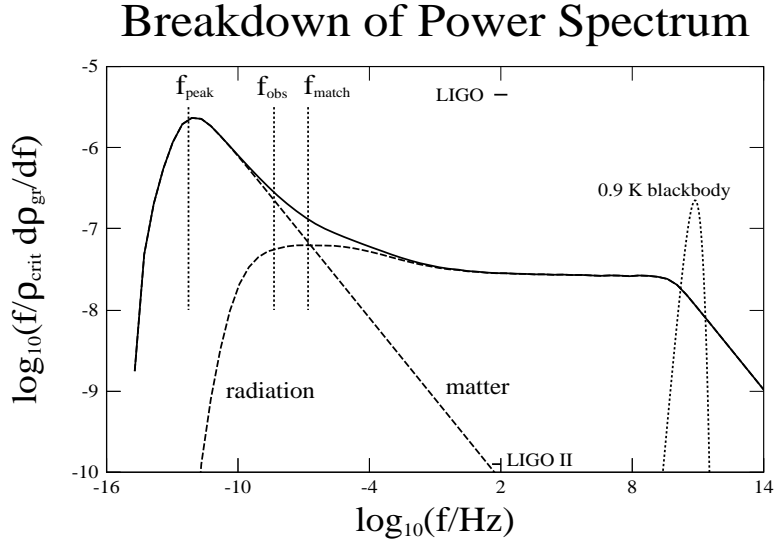, width=12cm, bbllx=0pt, bblly=0pt,
bburx=386pt, bbury=247pt}
\end{center}
\caption{ \label{f:cartoon}
A typical spectrum of gravitational radiation produced by a cosmic
string network is shown as the solid curve.  The dotted curves show the
contributions of loops which ``died" while the universe was
radiation-dominated, and the contributions of loops which ``died" more
recently, when the universe was matter-dominated.  The part of the
spectrum which lies in the LIGO band comes from the radiation-dominated
epoch.  Also shown, for naive comparison, is the spectrum of a 0.9
Kelvin blackbody.  The frequency $f_{\rm obs}$ is where the millisecond
pulsar timing observations constrain the spectrum.
}
\end{figure}

The spectrum of gravitational radiation produced by a cosmic string
network has been carefully studied
(Vilenkin 1985; Vilenkin \& Shellard 1994; Caldwell \& Allen 1992; 
Caldwell 1993).  The
spectrum of gravitational radiation produced by a ``typical" cosmic
string network is shown in Fig.~\ref{f:cartoon}.  You can see that
although the spectrum is not flat, it is fairly flat in the range of
frequencies which LIGO and the other ground-based detectors will be
sensitive to.  This radiation was produced by loops which emitted their
gravitational waves during the epoch when the universe was still
radiation-dominated.  I want to show you an estimate of the amplitude
of the stochastic spectrum $\Omega_{\rm gw}(f)$ for the range of
frequencies $f$ over which $\Omega_{\rm gw}$ is roughly constant.  To
do this, I need to carry out counting arguments, to count the number of
loops emitting gravitational radiation at any time. The number of loops
that we have depends upon the volume that we consider, and the counting
gets tricky if we don't make the right choice.  For example, we could
choose a volume which at any time was a single Hubble volume (a ball of
radius equal to the Hubble length), or we could pick a constant
co-moving volume, or a fixed physical volume.  To avoid the need to
make one of these choices, there is a simpler way to do the counting.
We'll do our calculations for a universe which has a finite spatial
volume, and whenever we count something (say loops), we'll simply count
the total number of those things (i.e. loops) in the {\it entire}
universe.

\begin{figure}
\centerline{\epsfig{file=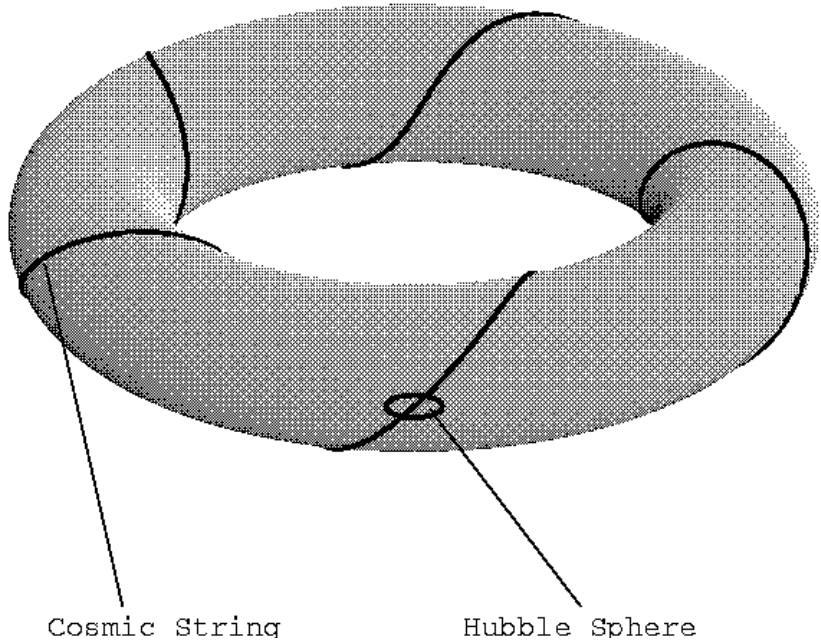, width=12cm, bbllx=100pt, bblly=220pt,
bburx=515pt, bbury=590pt}}
\caption{ \label{f:toroid}
A spatial slice of the model universe, at one instant in time, is a
three-dimensional torus.  In the figure, we have shown a single long
cosmic string, which winds around this universe 5 times in the
$x-$direction and 1 time in the $y-$direction.  Because the figure
shows a two-dimensional projection, we can't see how many times the
string winds around in the $z-$direction.   Note that the string has no
ends.  Also shown is a small sphere, which is the boundary of the
Hubble volume at the given instant in time.  At later times, this small
sphere would be larger, and would include a larger part of the
three-torus.  The single long string shown in this figure is a
so-called ``infinite" string.  Although its length is finite and it has
no ends, it is not called a ``loop" because it is longer than the
Hubble length.
}
\end{figure}

The universe we will use is shown in Fig.~\ref{f:toroid}.  At any
instant in time, its spatial topology is that of a three-dimensional
torus $T^3 = S^1 \times S^1 \times S^1$.  The metric is $ds^2=-c^2 dt^2 +
a^2(t)d\vec x^2$, but the spatial coordinates are each periodically
identified on the interval $[0,L]$.  Thus each of the three spatial
coordinates ranges over $0<x,y,z < L$.  The time coordinate $t$ is
physical or comoving time, and during the radiation-dominated epoch of
cosmological expansion, $a(t) = (t/t_0)^{1/2}$ where $t_0$ is the
present time, related to the present Hubble expansion rate by $t_0 = (2
H_0)^{-1}$.  Our model universe has a finite proper spatial volume; at
any instant in time this three volume is $V(t)=L^3 a^3(t)$.  Provided
that this volume is greater than the Hubble volume (in other words, $t$
is less than some critical value) then the fact that our model universe
is finite in spatial extent, rather than infinite, does not matter and
observable quantities will have the same values that they would have in
a truly infinite universe.  So the toroidal topology of our model
universe simply serves as a convenient calculational tool.

\begin{figure}
\begin{center}
\epsfig{file=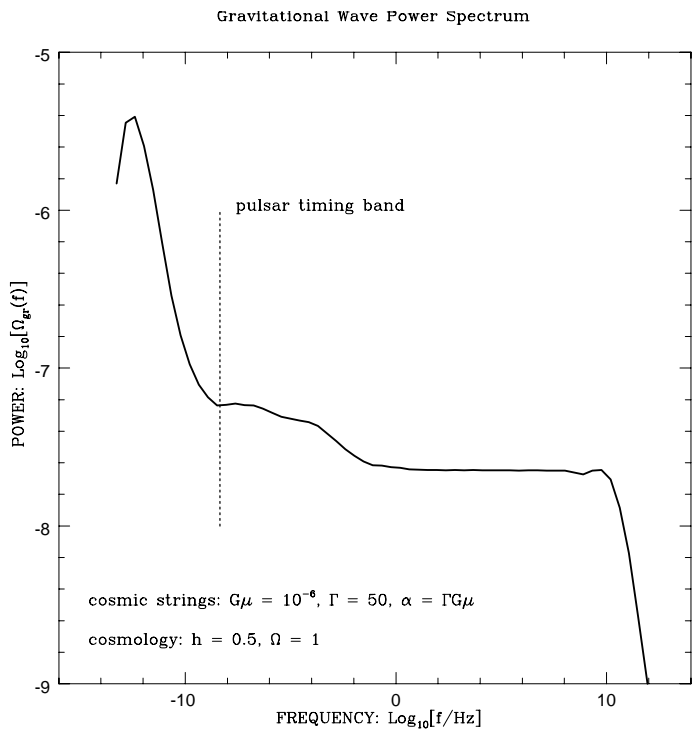, width=10cm, bbllx=18pt, bblly=154pt,
bburx=593pt, bbury=730pt}
\end{center}
\caption{ \label{f:stringspec}
The spectrum of gravitational radiation produced by a cosmic string
network.  Around the LIGO frequency band $f \approx 100$ Hz, the spectrum
is fairly flat.  In these lectures, we derive a simple analytic estimate
for the amplitude of $\Omega_{\rm gw}$ in this flat region.
}
\end{figure}

Shown in Fig.~\ref{f:stringspec}
is the spectrum of gravitational radiation produced by a cosmic string
network described by a particular set of dimensionless parameters.
This spectrum was actually calculated by a computer program
(Caldwell \& Allen 1992; Caldwell 1993); there is
no way to write a useful analytic expression for this spectra.  You
will notice however that the LIGO frequency band (around 100 Hz)
$\Omega_{\rm gw}$ is flat (a constant, independent of frequency).  It
is not difficult to derive a formula for this constant value.

To analytically determine the amplitude of $\Omega{\rm gw}$ in the LIGO
band, we will make extensive use of the scaling property of the string
network.  This property has been well established, both by computer
simulation and by analytic arguments 
(Vilenkin 1985; Vilenkin \& Shellard 1994).  The
scaling property may be neatly summarized:  {\it A single correlation
length $l(t)={\rm Hubble\ length}= 2c t$ characterizes ALL properties of
the string network.}  This scaling property can be used to obtain
qualitative relations.  For example, the energy-density of long strings is
given by
\be 
\rho_\infty = {A \mu l c^2 \over l^3} = {A \mu \over 4 t^2},
\ee
where $\mu$ is the mass-per-unit-length of the string, and $A$ is a
dimensionless constant (determined by numerical simulations to have a
value $A \approx 52$).  Roughly speaking, $A$ is the number of long
strings which pass through the Hubble volume shown in
Fig.~\ref{f:strings}. 
(An example of such a long string passing through a Hubble volume is
shown in Fig.~\ref{f:toroid}.)
Another example of scaling concerns the sizes of
loops of string which are ``chopped off" the infinite string network.
From scaling, the size of these loops at the time of formation must be
\be
{\rm Size\ of\ loop\ formed\ at\ }t_{\rm birth} = \alpha c t_{\rm birth},
\ee
where $\alpha$ is a dimensionless constant (which in principle can
be determined by numerical simulation).

Because the gravitational radiation produced by a cosmic string network
comes from the small oscillating loops of string, we need to determine
the rate at which these loops are formed.   This can be found from
energy conservation.  The decrease in the total energy of long strings
in the universe must be compensated by a corresponding increase in the
total energy of cosmic string loops:
\be
d(V \rho_\infty) + (\mu \alpha t c^3) d N_{\rm loops} = 0.
\ee
(Note that the equation of state of the string loops is similar to that
of dust; otherwise the equation of energy conservation would be
somewhat more complicated.) Using the expressions above for
$\rho_\infty$ and for $V$ we obtain an equation for the rate at which
loops of string are formed:
\be
\label{e:looprate}
{d N_{\rm loops} \over dt} = {1 \over 8} {L^3 A \over \alpha c^3 }
t_0^{-3/2} t^{-5/2},
\ee
where I've used the explicit time dependence of $a(t)$.
The loops which are being cut off the infinite strings oscillate
relativistically, emitting gravitational radiation.  A remarkable fact
about these loops is that the energy-loss rate (power radiated) $P$
does not depend upon the size of the loops, but only on their shape
(Vilenkin 1985; Vilenkin \& Shellard 1994).  For an ``average" loop
\be 
P = \gamma G \mu^2 c,
\ee
where $\gamma$ is a dimensionless constant, which has been determined
numerically to have a value $\gamma \approx 50$.
Because the loops radiate power at a constant rate, independent of their
size, the energy and mass of a loop decrease linearly with time, until
all the energy has been radiated away.  At time $t$, a loop formed at
time $t_{\rm birth}$ has length
\be
\left( {\rm Size\ of\ loop\ formed\ at\ }t_{\rm birth},{\rm \ at\ time\ }t
\right)  =
 \alpha c t_{\rm birth} -\gamma G \mu (t- t_{\rm birth})/c.
\ee
The loop disappears when this length reaches zero, at a time
\be
t_{\rm death} = \left(1 + {\alpha c^2 \over \gamma G \mu} \right)
t_{\rm birth} \equiv \beta t_{\rm birth},
\ee
where we have defined a constant $\beta = 1+ {\alpha c^2 \over \gamma G
\mu}>1$ which is the (constant) ratio of death-time to birth-time for
cosmic string loops.  We are now in a position to work out an analytic
approximation to $\Omega_{\rm gw}(f)$ for cosmic string networks, which
is valid in the LIGO band.

Consider the total energy (as seen today) emitted in gravitational
waves by all oscillating cosmic string loops which were formed (1)
after the cosmic string network came into existence at time $t_{\rm
formation}$ and (2) which are no longer present, and thus were born before
time $t_0/\beta$.  This total energy is
\bea
E_{\rm gw} &=& \int^{t_0/\beta}_{{t_{{\rm formation}}}}
dt' {d N_{\rm loop} \over dt'} \int_{t'}^{\beta t'} dt''
\gamma G \mu^2 c {a(t'') \over a(t_0)} \\
&&+ \bigl({\rm Ignore\ Loops\ Still\ Present\ Today} \bigr).
\eea
The different factors appearing in this formula are (1) the number of
loops $d N_{\rm loop}$ born in the time interval $dt'$, (2) the energy
$dE = dt'' \gamma G \mu^2 c$ emitted by those loops in the time
interval $dt''$, and (3) the redshift factor $a(t'') / a(t_0)$ which
reduces the energy as observed today.  Substituting in
expression (\ref{e:looprate}) for the rate of loop formation and
carrying out the integral over $t''$ gives
\bea
E_{\rm gw} &=& \int^{t_0/\beta}_{{t_{{\rm formation}}}}
dt' {d N_{\rm loop} \over dt'} 
\left[ {2 \over 3}  \gamma G \mu^2 c t_0^{-1/2} {t'}^{3/2}
\left( \beta^{3/2}-1 \right) \right]\\
&=&
{1 \over 8} {L^3 A \over \alpha c^2 } t_0^{-2} {2 \over 3} \gamma G \mu^2
\left( \beta^{3/2}-1 \right) \left[ \ln \left({t_0 \over t_{\rm formation}}
\right)
- \ln \beta \right].
\eea
Because of the scaling nature of the string network, 
and because frequencies redshift as $ t^{-1/2}$,
the logarithmic
frequency interval over which the gravitational radiation is
distributed is precisely
\be
\ln \left({f_{\rm high} \over f_{\rm low}}
\right)= {1 \over 2} \ln \left({t_0 \over t_{\rm formation}} \right).
\ee
The $\ln \beta$ term can be neglected in comparison with the other
$\ln$ term, because typically the magnitude of $100 < \beta < 10^5 $ is
much smaller than the magnitude of $t_0/t_{\rm formation} \approx
10^{50}$ for GUT scale strings.  If we now use the relationship between
the Hubble expansion rate and the critical energy density,
\be
H_0^2 = {1 \over 4 t_0^2} = {8 \pi G \rho_{\rm critical} \over 3 c^2}
\ee 
we can immediately calculate the spectrum of gravitational waves:
\bea
\Omega_{\rm gw} & = & {1 \over \rho_{\rm critical}}{d \rho_{\rm gw} \over d
\ln f} \\
& = &
{1 \over \rho_{\rm  critical}} {E_{\rm gw} \over \ln (f_{\rm
high}/f_{\rm low})} {1 \over L^3} \\ 
& = & {1 \over 8} {A \over \alpha c^2 } {32
\pi G \over 3 c^2} {2 \over 3} 2 \gamma G \mu^2 \left( \beta^{3/2}-1
\right).
\eea
Now this calculation is for a radiation-dominated universe.  If we include
the phase of matter domination following it, the amplitude of the
spectrum is decreased by a factor of $(1+Z_{\rm eq})$ to give
our final result for the amplitude of the stochastic gravity wave background
of a cosmic string network, in the LIGO band:
\be
\Omega_{\rm gw} = {16 \pi \over 9} {A \gamma \over \alpha}
\left( {G \mu \over c^2 } \right)^2 \left( \beta^{3/2}-1 \right)
(1+Z_{\rm eq})^{-1}.
\ee
This quantity can be large enough to be observable with the LIGO
detectors, for values of the dimensionless parameters characterizing
the cosmic strings.

The values of these dimensionless parameters are not completely known.
Reasonable values are $A=52$ for the number of long strings/horizon volume,
$\gamma \approx 50$ (for the radiation rate from a typical loop) and
mass-per-unit-length $G \mu / c^2 \approx 10^{-6}$ for GUT phase
transitions.  However the value of $\alpha$ (the size of a loop at
formation) is still unknown.  The high-resolution numerical simulations
have established that $\alpha<10^{-2}$ but this is only an upper
bound.  Indeed it is thought that $\alpha$ could perhaps be as small as
$\gamma G \mu/c^2 \approx 10^{-5}$, which is the scale at which
gravitational back reaction cuts off the small scale structure and
fluctuations on the long strings 
(Vilenkin 1985; Vilenkin \& Shellard 1994).  This
uncertainty in the value of $\alpha$ leads to a wide uncertainty in
$\beta$; the allowed range is $\beta \in [1,200]$.  Because this
parameter governs the lifetime of a loop, it has a large effect on the
spectrum $\Omega_{\rm gw}$.

There is increasing evidence that the scale of loop formation (i.e. the
value of $\alpha$) is indeed set by gravitational back reaction, and is
thus very small.   One can work out the limit $\alpha \to 0$ of
$\Omega_{\rm gw}$, obtaining \be \Omega_{\rm gw} = {8 \pi \over 3} A
\left( {G \mu \over c^2 } \right) (1+Z_{\rm eq})^{-1}.  \ee This may
well turn out to be the correct expression in the LIGO frequency
range.  However the allowable range of $\mu$ is tightly constrained by
the millisecond pulsar timing observations discussed in
Section~\ref{s:pulsar}, and the increasingly strong constraints of
these observations may rule out GUT scale cosmic strings in the near
future (Caldwell \& Allen 1992; Caldwell 1993).

Recent work \cite{martinvilenkin} has shown that ``hybrid" topological
defects could evade the increasingly strong constraints from pulsar
timing observations.  For example a cosmic string network might form at
early times, as described in this section, but then at later times,
domain walls form (having the strings as their boundaries) and ``pull
together" the strings, destroying the network.  In such a scenario
there is a high-frequency stochastic background (produced by the
strings at early times) but no low-frequency gravitational radiation.
In this way, the scenario evades the millisecond pulsar timing
constraints on $\Omega_{\rm gw}$ at $10^{-8}$ Hz.  The prospects for
detecting hybrid defects of this type with initial LIGO do not appear
good, however advanced LIGO should have more than enough sensitivity.

\subsection{Bubbles from first-order phase transitions}
\label{s:bubbles} 
My final example of a specific mechanism that can produce a stochastic
background of gravitational radiation is rather different than the
two previous examples I have examined (inflationary cosmology and
cosmic strings).  In these two previous examples there was a scaling
law or behavior which resulted in a spectrum $\Omega_{\rm gw}(f)$ that
was flat, at least for waves which entered the Hubble sphere before
the universe became matter-dominated.

\begin{figure}
\begin{center}
\epsfig{file=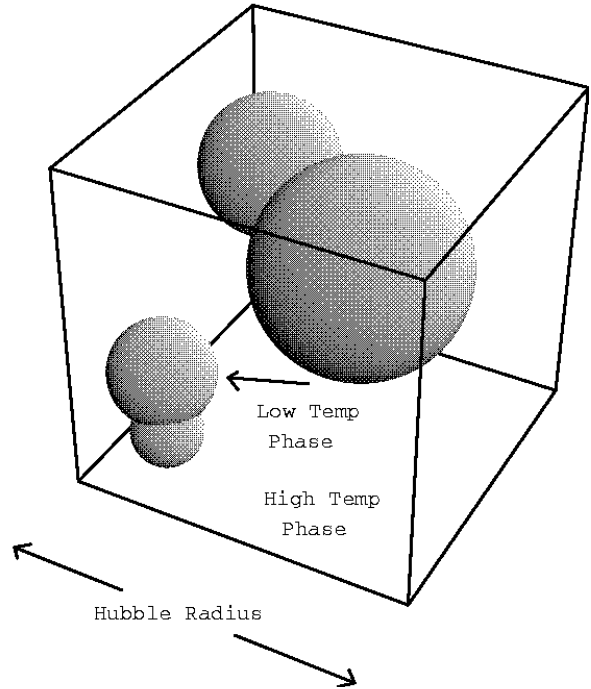, height=10cm, bbllx=125pt, bblly=165pt,
bburx=530pt, bbury=642pt}
\end{center}
\caption{ \label{f:bubbles}
A first-order phase transition can produce rapidly-expanding bubbles
containing the new (low temperature) phase within the old
(high-temperature) phase.  The difference in energy (latent heat) is
transformed into kinetic energy of the bubble walls.  When the bubbles
collide, a fraction of this energy is emitted as gravitational
radiation.
}
\end{figure}

In this final example, I will briefly review the effects of a different
type of mechanism.  In this mechanism, as the universe expands and
cools, the matter within it undergoes a first-order phase transition at
some definite time in the past 
(Kamionkowski {\sl et al.} 1994; Kosowsky {\sl et al.} 1992; 
Kosowsky \& Turner 1993).  This transition occurs
when the temperature $T_*$ of the universe has dropped sufficiently
below the characteristic energy of the phase transition: $kT_* < E_{\rm
transition}$.  Bubbles of the new (low-energy-density) phase form
within the old (high-energy-density) phase, as shown in
Fig.~\ref{f:bubbles}.
These bubbles expand rapidly after formation, converting the
difference in energy density $\times$ bubble volume into kinetic energy
of the bubble walls.  Within a short time, these bubbles are moving
relativistically, and within a Hubble expansion time, they collide.
These collisions are highly relativistic, non-symmetric events, and
produce copious amounts of gravitational radiation.  Within a short time
after the collisions begin, a significant fraction of the energy that
was in the bubble walls is converted into gravitational radiation.

\begin{figure}
\begin{center}
\epsfig{file=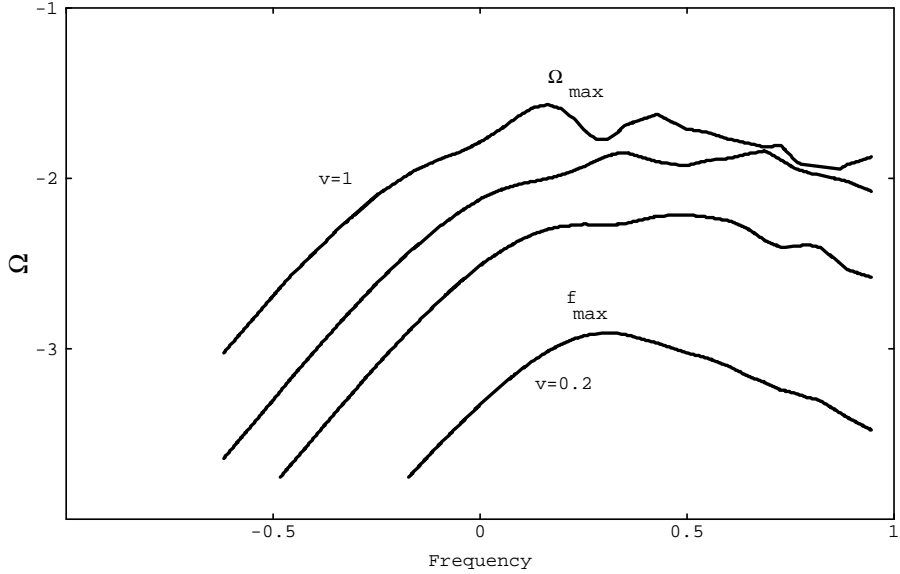, width=12cm, bbllx=72pt, bblly=245pt,
bburx=545pt, bbury=545pt}
\end{center}
\caption{ \label{f:bubblespect}
Typical spectra $\Omega_{\rm gw}(f)$ produced from the collision of
bubbles which result from a first-order phase transition.  This graph
is reproduced from Fig. 7 of Kamionkowski {\sl et al.} 1994.
The horizontal axis is $\log_{10}(2 \pi f/\beta)$ and the
vertical axis is $\log_{10} \Omega_{\rm gw} \beta^2 (1+ \alpha)^2
H^{-2} \alpha^{-2} \kappa^{-2}$.  The spectrum peaks at a
characteristic frequency $f_{\rm max}$, characteristic of the expansion
rate and time at which the bubble collisions occurred.  The spectra are
shown for bubble wall velocities of $v=0.2,0.4,0.6$ and $1.0$.}
\end{figure}

In contrast to our previous two examples, this bubble collision
process produces a spectrum which is strongly peaked at a particular
frequency, characteristic of the time at which the phase transition and
bubble collisions took place.  This spectrum has been computed
(Kamionkowski {\sl et al.} 1994; Kosowsky {\sl et al.} 1992; 
Kosowsky \& Turner 1993) and is shown in Fig.\ref{f:bubblespect}. 
The spectrum
is peaked at a characteristic frequency
\be
f_{\rm max} 
\approx 5.2 \times 10^{-8}
\left( {\beta \over H_*} \right)
\left({k T_* \over 1 {\rm \ GeV}} \right)
\left( {g_* \over 100} \right)^{1/6}
{\rm Hz}.
\ee
The Hubble expansion rate at the time of the phase transition is
denoted $H_*$ and $\beta$ is a measure of the bubble nucleation rate.
For typical phase transitions, Kosowsky, Turner and
their collaborators (Kamionkowski {\sl et al.} 1994; Kosowsky {\sl et al.} 1992; 
Kosowsky \& Turner 1993) find that $\beta/H_* \approx 4
\ln(M_{\rm Planck} c^2 /k T_*) \approx 10^2$.  The number of degrees of
freedom $g_*$ is also of order $10^2$ in typical GUT models.  For
example, in the minimal standard model electro-weak phase transition
one has the transition at an energy $k T_* \approx 10^2 \rm \ GeV$ with
$f_{\rm max} \approx 4.1 \times 10^{-3} \rm \ Hz.$

The amplitude of the spectrum depends mostly upon the relative
energy-density difference between the interior and the exterior of a
bubble.  This difference in energy density is deposited directly in the
bubble walls, determining the speed at which the walls crash together,
and thus the amount of energy converted to gravitational radiation.
The difference is described by a parameter $\alpha={\rho_{\rm vac}
\over \rho_{\rm thermal}}$.  Kosowsky, Turner and Watkins find that the
amplitude of the spectrum at the peak is approximately
\be
\Omega_{\rm gw}(f_{\rm max}) h_{100}^2 \approx 1.1 \times 10^{-6}
\kappa^2 \left( {\beta \over H_*} \right)^{-2} \left( {\alpha \over 1 +
\alpha} \right)^2 \left( {v^3 \over 0.24 + v^3 } \right) \left( {g_*
\over 100} \right)^{-1/3}.
\ee
The parameter $\kappa$ is the fraction of vacuum energy that goes into
kinetic energy of the fluid; it is an increasing function of $\alpha$
and typically lies in the range $10^{-2} < \kappa < 1$.  The final
parameter $v$ is the propagation velocity of the bubble walls.  For
strongly first-order phase transitions, $\alpha \to \infty$ and $v \to
1$.  In this strongly first-order case, a phase transition at $\approx
10^9 \> \rm GeV$ could be easily observed with the Advanced LIGO
detectors.

We should note that it is perhaps overly optimistic to assume that
the phase transition is
strongly first-order.  At least one phase
transition which is expected to occur, at an energy scale of about
$10^2$ GeV, is the electro-weak transition.  In the standard SU(3)
model this phase transition has $\Omega_{\rm gw}(4.1 \times 10^{-3} \> \rm
Hz) \approx10^{-22}$, which is many orders of magnitude too small to be
observable by any proposed space-based detector 
(Kamionkowski {\sl et al.} 1994; Kosowsky {\sl et al.} 1992; 
Kosowsky \& Turner 1993).
However it is perfectly possible that there as a strongly first-order
phase transition in the earlier history of the universe, that might
give rise to waves observable with the proposed LISA detector.

\section{Conclusion}
\setcounter{equation}0
\label{s:conc} 
I am often asked, after a series of lectures like this, ``Do you really
think that a stochastic background is there?" There is only one
possible honest answer -- I do not know!  The problem is that we have
very little information about what really happened in the very early
universe.  And the scenarios that we have given as examples are just
that, examples.  Personally, I think it is unlikely that any of them
will turn out to be true.  However they do have {\it illustrative}
value.  They show that, under reasonable assumptions about what
happened in the early universe, it {\it is} possible to produce a
stochastic background which is easily detected with the instruments now
under construction.  The possibility of such detection has
been anticipated for some time (Grishchuk 1976; 1988), and there
are good hopes that they will be realized in the next decade.

So while the answer to the question is unknown, that does not matter.
In the next decade we will probably know the answer in a very direct
way, with real observational data.  And if we do detect a stochastic
background, a new era in our understanding of the early universe will
begin.

\begin{acknowledgments}
This work has been partially supported by NSF grant PHY95-07740.
I also thank Kip Thorne for his hospitality at Caltech, where much of
this work was completed, and for the support of his NSF grant
PHY94-24337.
I acknowledge
B.S. Sathyaprakash for showing me the
elegant argument which I have adapted at the end of 
Section~\ref{s:optimal},
and useful conversations with Eanna Flanagan, Kip Thorne, Rai Weiss, 
Kent Blackburn, David Shoemaker, Ron Drever, Stan Whitcomb, and Robbie Vogt.
I am also very grateful to Joseph Romano for his careful reading of this manuscript
and many useful comments, and to Alex Vilenkin for pointing out an
error in the original manuscript.
\end{acknowledgments}

\appendix
\section{Calculational details}
\setcounter{equation}0

\label{s:appendix1} 
The purpose of this section is to provide details of the calculations
of the expectation values of the strains due to a stochastic background.
The starting point is the plane wave expansion of the gravitational
metric perturbations.  In transverse traceless gauge, this can be
written in the form of a plane wave expansion
\be
h_{ab}(t,\vec x) = \sum_A \int d^3 k \> C_A(\vec k) \sin\bigl(c |\vec k| t - \vec k
\cdot \vec x + \Phi(\vec k) \bigr) e_{ab}^A( \hat k).
\ee
Here $A=+,\times$ label the two polarization states, $\vec k$ is the
wave-vector of the perturbation, $C_A$ are arbitrary real functions
specifying the amplitude of the modes, $\Phi$ is a real function taking
values in the interval $[0,\pi]$ and $e_{ab}^A$ are the polarization
tensors, which depend only upon the direction of the wave-vector $\vec
k$.  This decomposition is unique:  every gravity wave can be
represented exactly one way, in this form.

Rather than representing the traveling wave by a phase and a real
amplitude, it is more convenient to use a complex amplitude.  An
equivalent representation is
\be
h_{ab}(t,\vec x) = \Re \left( \sum_A \int_0^\infty df \int_{S^2} d\hat
\Omega \> B_A(f,\hat \Omega) \exp(2 \pi i f(t-\hat \Omega \cdot \vec x/c))
e_{ab}^A(\hat \Omega) \right)
\ee
Here, $\hat \Omega$ is a unit vector on the two-sphere; we have
effectively written $\vec k = 2 \pi f \hat \Omega$.  The function
$B_A(f,\hat \Omega)$ is an arbitrary complex function of the two
variables, defined on the range $f \ge 0$.  By taking the real part of
this expression explicitly, one can cast it in the form
\be
h_{ab}(t,\vec x) =  \sum_A \int_{-\infty}^\infty df \int_{S^2} d\hat
\Omega \> h_A(f,\hat \Omega) \exp(2 \pi i f(t-\hat \Omega \cdot \vec x/c))
e_{ab}^A(\hat \Omega),
\ee
where now $h_A(f,\hat \Omega)$ is an arbitrary complex function
satisfying the relation $h_A(-f,\hat \Omega) = h_A^*(f,\hat \Omega)$.
As before, this decomposition is unique.

The polarization tensors appearing in these relations may be given
explicitly.   In standard angular coordinates $(\theta,\phi)$ on the
sphere one has
\bea
\hat \Omega & = & \cos \phi \sin \theta \hat x + \sin \phi \sin \theta \hat y
+ \cos \theta \hat z \\
\hat m& = & \sin \phi \hat x - \cos \phi \hat y \\
\hat n& = & \cos \phi \cos \theta \hat x + \sin \phi \cos \theta \hat y 
- \sin \theta \hat z \\
\label{e:polar}
e_{ab}^+(\hat \Omega) & = & m_a m_b - n_a n_b \\
\label{e:polar2}
e_{ab}^\times(\hat \Omega) & = & m_a n_b + n_a m_b
\eea
One can verify by inspection that $\hat m$ and $\hat n$ are a pair of
orthogonal unit-length vectors in the plane perpendicular to $\hat
\Omega$.  It is simple to show that any rotation of the vectors $\hat
m$ and $\hat n$ within the plane that they define simply corresponds to
a trivial re-definition of the complex wave amplitudes $h_+$ and
$h_\times$.

We are now in a position to verify some of our key formula.  We begin
with the assumption that the stochastic background is stationary and
Gaussian.  This means that the ensemble average of the Fourier
amplitudes is
\be
\langle h_A^*(f, \hat \Omega) h_{A'}(f',\hat \Omega') \rangle =
  \delta(f-f') \delta^2(\hat \Omega ,\hat \Omega') \delta_{AA'} H(f)
\ee
for all $f$ and $f'$.  In these
formula, 
\be \delta^2(\hat \Omega,\hat \Omega') = 
\delta(\phi-\phi') \delta(\cos \theta - \cos \theta')
\ee
is the covariant Dirac $\delta$-function on the two-sphere, and the
covariant volume element is $ d \hat \Omega = \sin \theta d\theta d\phi
$.  The function $H(f)$ is related to the spectrum $\Omega_{\rm
gw}(f)$.  $H(f)$ is a real, non-negative function, satisfying
$H(f)=H(-f)$.
To see the relationship, consider the energy
density in gravitational waves.  This is given (locally) by
\be \rho_{\rm gw} = {c^2 \over 32 \pi G} \langle \dot h_{ab} \dot h^{ab} \rangle,
\ee
where the overdot denotes a time derivative, and both tensors are
evaluated at the same space-time point $(t,\vec x)$.
Substituting the plane wave expansion into this formula yields
\be 
 \langle \dot h_{ab}(t,\vec x) \dot h^{ab}(t,\vec x) \rangle
= \sum_A \int_{-\infty}^\infty df \int_{S^2} d \hat \Omega \>
4 \pi^2 f^2 H(f) e^A_{ab}(\hat \Omega) e_A^{ab}(\hat \Omega).
\ee
Since $\sum_A e_{ab}^A e^{ab}_A = 4$ and $\int d\hat \Omega = 4 \pi$
one has
\be \label{e:rhoingw}
\langle \dot h_{ab} \dot h^{ab} \rangle=64 \pi^3 \int_{-\infty}^\infty df \> f^2 H(f)= 128 \pi^3 \int_0^\infty df \> f^2 H(f).
\ee
Using the definition $\Omega_{\rm gw}$ one obtains the
relationship between the spectrum $\Omega_{\rm gw}$ and $H(f)$.
For $f \ge 0$ one has
\be \Omega_{\rm gw}(f)  = {f \over \rho_{\rm critical}} {d \rho_{\rm gw} \over df}
= f { 8\pi G\over 3 c^2 H_0^2} {c^2 \over 32 \pi G} 128\pi^3 f^2 H(f)
= {32 \pi^3 \over 3 H_0^2} f^3 H(f).
\ee
This formula is needed to obtain the next result.

The second relation required is the expected value of the Fourier amplitudes
at two different sites.  This is now easily obtained.  By definition
\beau
& & \langle \tilde h_1^*(f) \tilde h_2(f') \rangle \\
&=& \sum_{AA'} \int_{S^2} d\hat \Omega \int_{S^2} d\hat \Omega'
\langle h_A^*(f,\hat \Omega) h_{A'}(f',\hat \Omega') \rangle \\
& & \quad \times \exp
\bigl( 2 \pi i f \hat \Omega \cdot \vec x_1/c -
 2 \pi i f' \hat \Omega' \cdot \vec x_2/c \bigr) F_1^A(\hat \Omega)
F_2^{A'}(\hat \Omega') \\
& = & 
\sum_{A} \int_{S^2} d\hat \Omega H(f) \exp \bigl( 2 \pi i f\hat \Omega
\cdot \Delta \vec x/c \bigr) F_1^A(\hat \Omega) F_2^{A}(\hat \Omega) 
\delta(f-f')
\\
& = & 
{8 \pi \over 5} \gamma(f) H(f) \delta(f-f') = {3 H_0^2 \over  20 \pi^2} |f|^{-3} \Omega_{\rm
gw}(|f|) \gamma(|f|) \delta(f-f').
\eeau
In the third line above, we have used the definition of the strain in
the detector, given in (\ref{e:detectresponse}), and in the fourth
line, we have used the definition of the overlap reduction function
(\ref{e:overlap}).

%
%

\end{document}